\documentclass[10pt]{iopart} 
\usepackage{amsmath}
\usepackage{amsmath}
\usepackage{amssymb}
\usepackage{graphicx}
\usepackage{color}
\usepackage{accents}

\newcommand{\erf}{\mbox{erf}}
\renewcommand{\e}{\epsilon}

\newcommand{\beq}{\begin{equation}}
\newcommand{\eeq}{\end{equation}}
\newcommand{\be}{\begin{equation}}
\newcommand{\ee}{\end{equation}}
\newcommand{\beqn}{\begin{eqnarray}}
\newcommand{\eeqn}{\end{eqnarray}}
\newcommand{\bea}{\begin{eqnarray}}
\newcommand{\eea}{\end{eqnarray}}
\newcommand{\bearr}{\begin{array}}
\newcommand{\enarr}{\end{array}}

\renewcommand{\vec}{\tilde}

\newcommand{\commento}[1]{}

\begin{document} 

\title{Onsager coefficients in a coupled-transport
model displaying a condensation transition}

\author{Stefano Iubini$^{1,2,*}$, Antonio Politi$^{3,1}$, Paolo Politi$^{1,2}$}

\address{$^{1}$ Istituto dei Sistemi Complessi, Consiglio Nazionale delle Ricerche, via Madonna del Piano 10, I-50019 Sesto Fiorentino, Italy}
\address{$^{2}$ Istituto Nazionale di Fisica Nucleare, Sezione di Firenze, via G. Sansone 1 I-50019, Sesto Fiorentino, Italy}
\address{$^{3}$ Institute for Complex Systems and Mathematical Biology, 
           University of Aberdeen, Aberdeen AB24 3UE, United Kingdom}
\address{$^{*}$ Author to whom any correspondence should be addressed}
\eads{\mailto{a.politi@abdn.ac.uk}, \mailto{paolo.politi@cnr.it}, \mailto{stefano.iubini@cnr.it}}

\begin{abstract}
We study nonequilibrium steady states of a one-dimensional stochastic model, originally introduced as an
approximation of the Discrete Nonlinear Schr\"odinger  equation.
This model is characterized by two conserved quantities, namely  mass and energy; it
displays a ``normal", homogeneous phase, separated by a condensed (negative-temperature) phase, 
where a macroscopic fraction of energy is localized on a single lattice site.
When steadily maintained out of equilibrium by external reservoirs, the system exhibits coupled transport
herein studied within the framework of linear response theory.
We find that the Onsager coefficients satisfy an exact scaling relationship, which allows reducing
their dependence on the thermodynamic variables to that on the energy density for unitary mass density.
We also determine the structure of the nonequilibrium steady states in proximity of the critical line,
proving the existence of paths which partially enter the condensed region.
This phenomenon is a consequence of the Joule effect:
the temperature increase induced by the mass current
is so strong as to drive the system to negative temperatures.
Finally, since the model attains a diverging temperature at finite energy, in such a limit 
the energy-mass conversion efficiency reaches the ideal Carnot value.
\end{abstract}


\maketitle

\section{Introduction}

The characterization of non-equilibrium steady states (NESS) is an important research area, as
Nature is plenty of systems that steadily exchange physical quantities (e.g., energy, mass) with the surrounding environment. 
In this area, linear response theory represents a cornerstone; it allows, in fact, expressing
transport coefficients in terms of equilibrium fluctuations, under the condition of weak currents.
The treatment of systems far from equilibrium is still a challenge although some progress has been
recently made thanks to the application of large-deviation theories~\cite{Sornette_book,touchette09}.

Anyway, even in the realm of regimes close to equilibrium, there are nontrivial open questions.
For instance, in low-dimensional systems, the long-range correlations which characterize NESS 
may be so important as to induce anomalous (diverging) conductivity~\cite{Lepri_PhysRep,lepri2016,livi22}. 
In this context, additional tools based on fluctuating hydrodynamics~\cite{Spohn23} are required to account for the
resulting scenario, and yet a few challenging exceptions still need be explained (see the summary in~\cite{lepri20}).

Another area concerns coupled transport phenomena in systems where two or more quantities are 
simultaneously transported. For instance, identifying the conditions of an optimal thermoelectric
conversion is very important because of potential applications for energy production and storage~\cite{benenti17rev}. 
A general answer will likely require substantial progress on the basic mechanisms; for instance,
in Ref.~\cite{benenti14} it has been discovered that combining coupled transport with anomalous
transport may be a way to increase the efficiency.

Altogether, the analysis of simple models is potentially very useful because they allow both
for the development of analytical treatments and for the performance of detailed numerical investigations. 
In fact, the one-dimensional setup, where a chain of point-like elements is put in contact with two different 
reservoirs at its ends, has attracted much attention~\cite{Lepri_PhysRep}.
The most paradigmatic example is perhaps the asymmetric simple exclusion process (ASEP)~\cite{livi17}, introduced to describe
the transport of particles through a channel, which revealed quite useful to describe molecular motors~\cite{Klumpp2003}.
Another celebrated example is the exactly solvable model proposed in 1982 by Kipnis, Marchioro and Presutti
to describe energy diffusion in one-dimensional systems~\cite{kmp82}. Here, the presence of long-range correlations is explicitly accounted for
and the invariant non-equilibrium measure is exactly known.
Linear oscillators accompanied by random elastic collisions are another enlightening model, where the stochastic collisions
 mimic nonlinearities, destroying the integrability of the harmonic chain. 
This model is indeed an archetypical system displaying anomalous heat conductivity~\cite{basile06}.

The above  methodology turns out to be  useful also in the context of coupled transport problems, as
very little is known on the thermodiffusive behavior of interacting systems from the statistical mechanical point of view~\cite{benenti17rev,maes05,Iubini2016_NJP}.

In this paper, we focus on a stochastic interacting system, which, roughly speaking, extends the model proposed and solved 
in Ref.~\cite{kmp82}.
Here, there is a single set of microscopic-state variables $c_i$, and two positive-defined conserved quantities, 
formally identifiable with total mass and energy. 
This model can be seen as a simplified stochastic version of the Discrete NonLinear Schr\"odinger (DNLS) equation,
which arises ubiquitously in nonlinear physics and displays important applications in cold-atoms physics and  nonlinear optics~\cite{kevrekidis09}. 
In this perspective, the variable $c_i$ has the physical meaning of the local norm of the DNLS wavefunction on lattice
site (see~\cite{Iubini2013_NJP,JSP_DNLS,JSTAT_mmc,Barre2018_JSTAT,arezzo22}
for related studies and details on the derivation). 

The model was originally introduced to investigate the
spontaneous onset of energy localization in the DNLS  dynamics~\cite{Iubini2013_NJP}, 
but it has become a typical example of constraint-driven condensation~\cite{Szavits2014_PRL,Szavits2014_JPA,GIP21}.
In fact, depending on the densities of the two conserved quantities, a finite fraction of energy may eventually condense
on a single site, a regime characterized by a negative absolute temperature~\cite{Gradenigo2021_JSTAT,Baldovin_2021_review}. 
At equilibrium, a critical line of infinite-temperature states separates the condensed region from 
a homogeneous one displaying standard equipartition~\cite{Gradenigo2021_EPJE}.
In this model, the energy localization mechanism is the direct consequence of the existence of two 
conservation laws along with the positivity constraint, $c_i\ge 0$.
For this reason, we refer to it as C2C (condensation with two conserved quantities);
see the next section for a more precise definition.

When a C2C chain is attached to two different reservoirs at its boundaries, the corresponding NESS can be visualized as a path in the
thermodynamic plane (characterized by either
mass and energy density in the microcanonical plane, or chemical potential and temperature in the grand canonical plane).
Recently, it has been found that such paths may spontaneously enter the condensed region
even when the extrema lie in the homogeneous region~\cite{gotti22}\footnote{A similar scenario was previously observed 
also in simulations of the DNLS equation with a pure dissipation acting on one edge and a
standard reservoir on the other side~\cite{Iubini2017_Entropy}.}.
In practice, enegy can robustly localize within an internal portion of the chain, while the boundaries of the system
behave smoothly and display standard thermal fluctuations.
These examples reveal a novel type of condensation that takes place exclusively in out-of-equilibrium conditions and
in the presence of coupled transport.

The accessibility of such unusual nonequilibrium states is manifestly a relevant research subject in the context of
irreversible thermodynamics.
Here, we thoroughly explore the NESS emerging in the C2C model with the help of linear response theory,
devoting a special attention to the behavior close to the critical line.
An exact scaling analysis shows that the two-parameter dependence of the thermodynamic states
can be reduced to the dependence on a single variable.
This includes the coefficients of the Onsager matrix, whose behavior is crucial to reconstruct NESS paths.

We find that the Onsager coefficients are 
well-defined and finite along the whole critical line of the model, thus making it possible a
novel and potentially useful
kind of ``infinite-temperature 
transport''. 
We derive perturbative expressions of the corresponding paths and, more important,  we identify
the class of paths which are due to enter the condensed phase. Qualitatively, this phenomenon can be
seen as an extreme version of a Joule effect: the mass current induced by the external reservoirs heats up
the interior of the system forcing it to go beyond the infinite-temperature line, i.e. to condense.
Moreover, we show that in proximity to the critical line,
the thermodiffusive conversion performance reaches its maximum value
(as expressed in terms of the Carnot efficiency).

We are able to show that nonequilibrium correlations arising in NESSs play an important role in the high-temperature limit,
thus making the C2C transport substantially different from that of noninteracting dilute gases.
Approximate analytic expressions of the Onsager transport coefficients are obtained
in the opposite regime of small-temperatures, where spatial correlations are negligible.
There, we also find that the Seebeck coefficient,
which quantifies the coupling between mass and energy currents, vanishes rather rapidly when the ground state is approached.

The paper is organized as follows.
In Sec.~\ref{sec:model} we introduce the model, its equilibrium and out-of-equilibrium properties,
and show that (local) equilibrium properties depend on one parameter only: a suitable combination
of mass and energy in the microcanonical ensemble; a combination of temperature and
chemical potential in the grand canonical ensemble.
In Sec.~\ref{sec:Onsager} we introduce and determine the Onsager coefficients.
Because of scaling properties, it is sufficient to determine them for a unitary value of the mass density.
In Sec.~\ref{sec:corr} we evaluate spatial correlations and discuss their increasing importance when temperature grows.
In Sec.~\ref{sec:tn} we estimate the steady-state paths in proximity of the critical line and
determine the condition for them to enter the
negative-temperature region.
In Sec.~\ref{sec:conversion} we discuss the Seebeck coefficient and the conversion efficiency.
Finally, in Sec.~\ref{sec:conc} we provide some concluding remarks.
\ref{app:pert} contains some technical details, and a slightly different model is presented in \ref{sec:nopin}.

\section{The model and its equilibrium and out-of-equilibrium properties}
\label{sec:model}

The C2C model is defined on a lattice whose nodes $i$ host a non negative quantity
$c_i \ge 0$, here called (local) mass. Its square is called (local) energy, $\e_i \equiv c_i^2$.
Both the total mass $(A=\sum_i c_i)$ and the total energy $(H=\sum_i \e_i)$ are conserved
and the model is microcanonically defined through the mass density $a=A/N$ and the
energy density $h=H/N$, where $N$ is the total number of sites.

The equilibrium properties are well understood~\cite{Szavits2014_PRL,Gradenigo2021_JSTAT,Gradenigo2021_EPJE,gotti22}. 
The system has a homogeneous phase for $a^2 \le h \le 2a^2$ and a condensed/localized phase
for $h>2a^2$, characterized in the thermodynamic limit by a single site hosting a finite
fraction of the whole energy, equal to $(h-2a^2)N$. Finite-size effects provide an interesting
and unexpected scenario close to the critical line, $h_c=2a^2$~\cite{GIP21}.
When $h$ varies between the ground state $h=a^2$ and the upper value of the homogeneous phase, $h=h_c$,
the temperature varies between $T=0$ and $T=+\infty$.
In the localized region, the temperature is constant and equal to $T=+\infty$ but subleading terms
in the entropy (i.e., non extensive terms) show that finite-size systems are characterized by
a \textit{negative} temperature when $h_c < h < h_c + \xi(N)$, where $\xi(N) \simeq 11.05/N^{1/3}$, see Ref.~\cite{Gradenigo2021_JSTAT}.
In such a region the system is effectively delocalized~\cite{Gradenigo2021_EPJE}.

The grand canonical description is well defined in the homogeneous phase only, $h\le h_c$,
where there is ensemble equivalence~\cite{Gradenigo2021_JSTAT}.
The grand canonical partition function reads
\be
Z(\beta,\mu) = \left[\int_0^{+\infty}dc\, e^{-\beta( c^2 -\mu c)}\right]^N\, .
\label{eq:Z}
\ee
As already detailed in Ref.~\cite{gotti22},
the inverse temperature $\beta=1/T$ and the chemical potential $\mu$ are related to the
mass density $a$ and to the energy density $h$ through the relations
\bea
\label{eq.a_gc}
a &=& \frac{\mu}{2} + \frac{1}{\sqrt{\pi\beta}}
\frac{e^{-\beta\mu^2/4}}{1 +\erf\left(\frac{\sqrt{\beta}\mu}{2}\right)}, \\
h &=& \frac{1}{2\beta} +\frac{1}{2}a\mu .
\label{eq.h_gc} 
\eea

Eqs.~(\ref{eq.a_gc},\ref{eq.h_gc}) provide a mapping from grand canonical to microcanonical quantities.
While numerical simulations allow for a direct extraction of $(a,h)$, the Onsager coefficients are
better expressed in terms of grand canonical quantities.
It is therefore useful to invert the above mapping. We start introducing
\be
m = \mu \beta .
\label{eq:m}
\ee
as this also helps unveiling a general scaling dependence.
If we further introduce the auxiliary variable
\be
z = \frac{m}{\sqrt{\beta}} ,
\label{eq:z}
\ee
relations (\ref{eq.a_gc}-\ref{eq.h_gc}) can be rewritten as
\begin{eqnarray}
a &=& \frac{1}{m}\left[ 
\frac{z^2}{2} + \frac{z}{\sqrt{\pi}} \frac{\mathrm{e}^{-z^2/4}}{1 + \mathrm{erf}(z/2) }\right] 
\equiv \frac{\tilde m(z)}{m} \label{eq:orig1} \\
h &=& a^2 \cdot \frac{z^2}{2} \left( \frac{1}{\tilde m^2(z)} + \frac{1}{\tilde m(z)} \right) \equiv a^2 {\vec h}(z) . \label{eq:orig2}
\end{eqnarray}
Eq.~(\ref{eq:orig2}) demonstrates that the specific energy $\vec h \equiv h/a^2$ is a function of $z$ only.
Indeed, the invariance of the model under a uniform rescaling of the $c_i$'s implies that
the mass density $a$ is essentially a unit of measure and that
global and local equilibrium properties depend only on $\vec h$, or equivalently on $z$.

\begin{figure}
\begin{center}
\includegraphics[width=0.7\textwidth,clip]{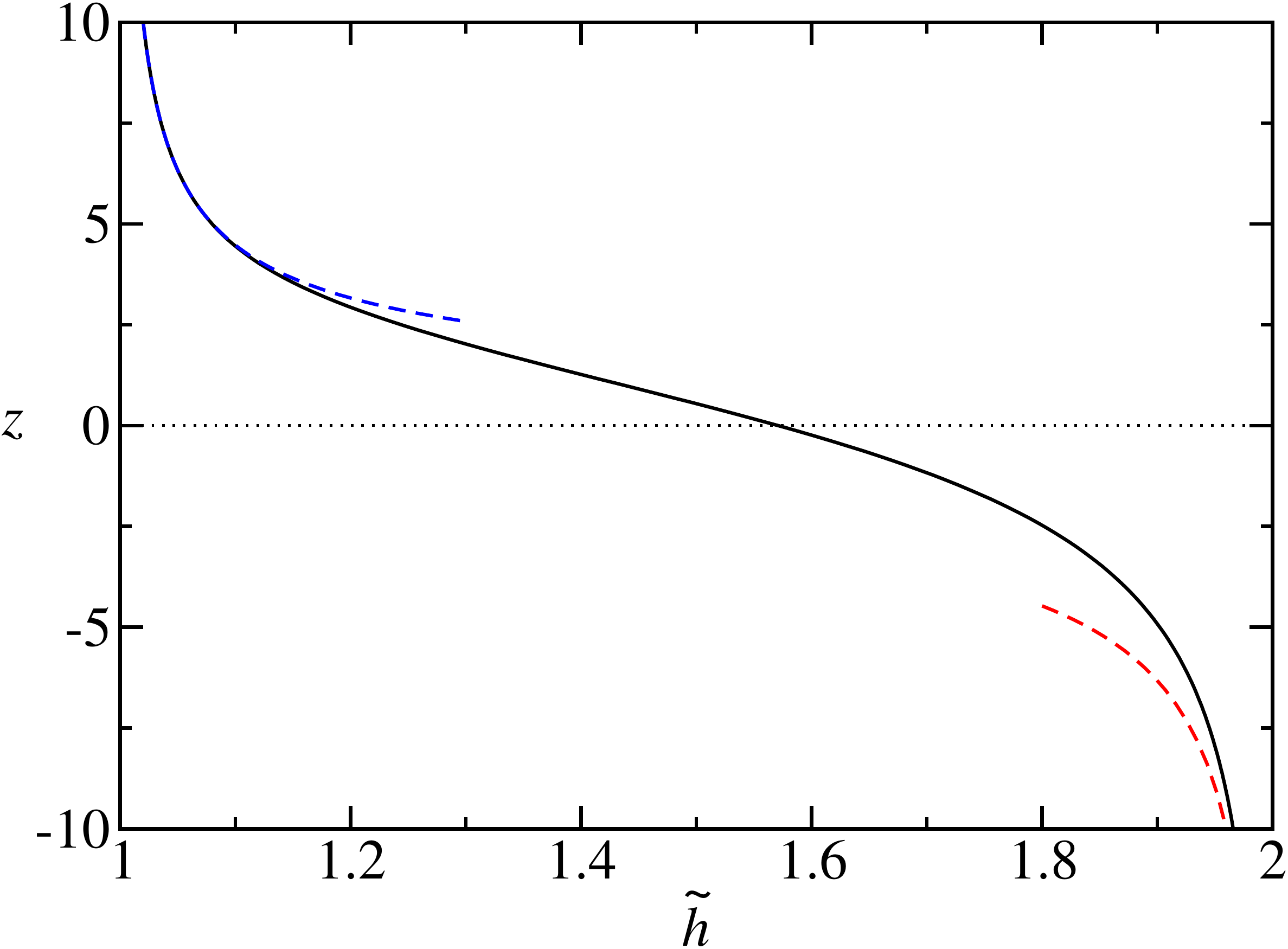}
\caption{The dependence of $z$ on the effective energy $\vec h$ as determined by
Eq.~(\ref{eq:orig2}). As $\vec h$ varies from $1$ to $2$, the temperature increases from $0$
to $+\infty$, and $z=\mu\sqrt{\beta}$ varies between $+\infty$ and $-\infty$. The change of
sign of $z$ corresponds to a change of sign of the chemical potential.
The dashed lines represent the analytical approximations of Eq.~(\ref{eq:zan}),
valid for vanishing and diverging temperature.}
\label{fig:hz}
\end{center}
\end{figure}

In fact, $\vec h(z)$ plays a crucial role to understand the relastionship between the microcanonical and grand canonical
representention. First of all, as shown in Fig.~\ref{fig:hz}, this function is one-to-one and thus perfectly
invertible.
In the same figure we plot also the limiting behavior determined via a perturbative analysis carried out in~\ref{app:pert},
\bea
z \simeq \left\{
\begin{array}{lcc}
\displaystyle \sqrt{2/(\vec h -1)} & & \vec h \to 1 \\
\displaystyle -2/\sqrt{2 -\vec h} & & \vec h \to 2 . 
\end{array}
\right.
\label{eq:zan}
\eea

\begin{figure}
\begin{center}
\includegraphics[width=0.7\textwidth,clip]{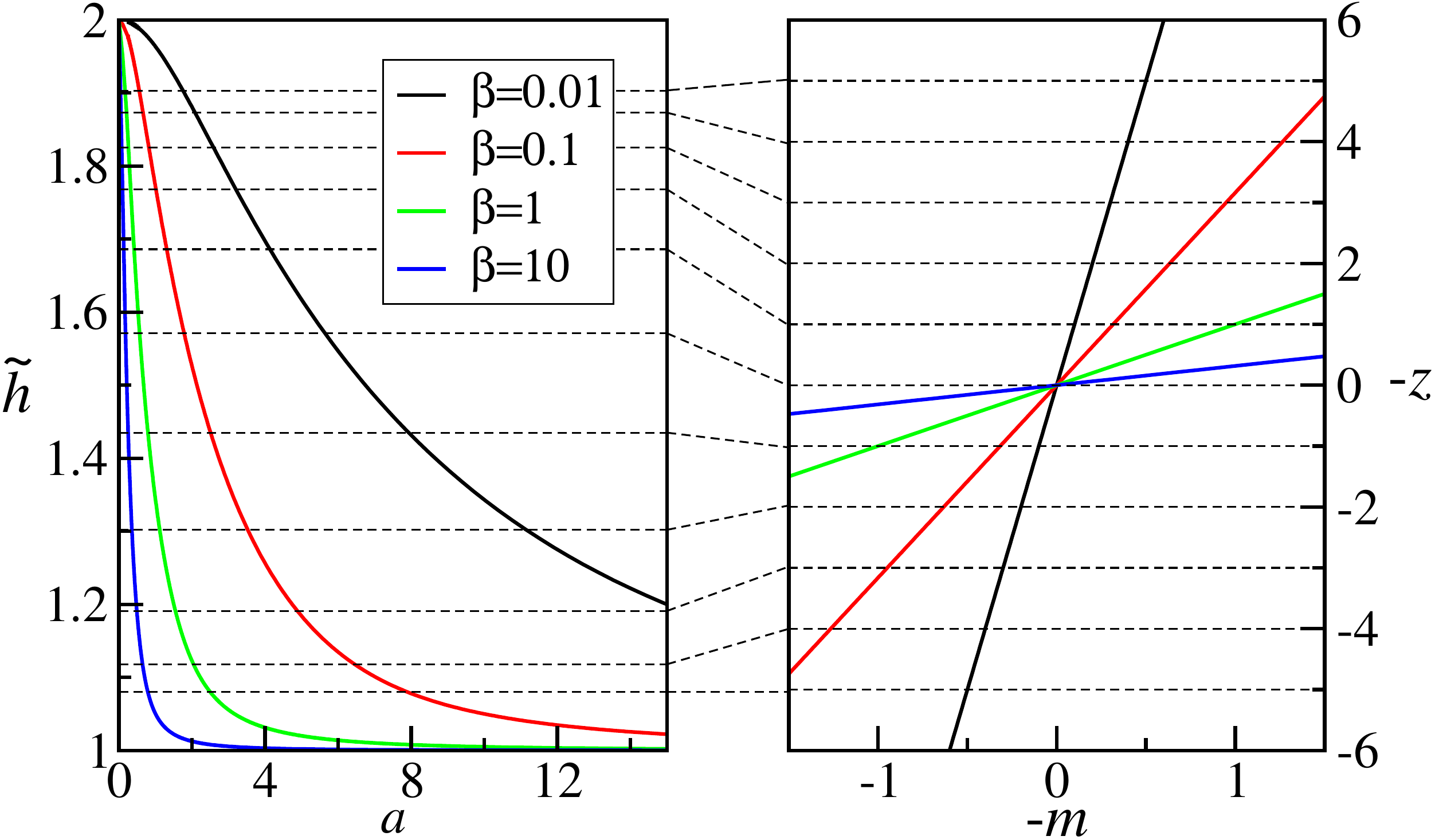}
\caption{Isothermal lines $\beta=$ const in the $(a,\tilde h)$ space (left panel) and in the $(m,z)$ one (right panel).
Dashed lines are examples of curves at constant $\tilde h$ which are mapped to horizontal lines $z=$ const by Eqs.~(\ref{eq:orig1}-\ref{eq:orig2}). }
\label{fig:curves}
\end{center}
\end{figure}

In practice, from the $a$ and $h$ values, the specific energy $\vec h$ is first obtained and thereby the corresponding $z$ value.
Afterwards $m$ can be obtained as $\tilde m(z)/a$ and finally $\beta = m^2/z^2$.
A schematic representation of the mapping is presented in Fig.~\ref{fig:curves}.

For $\tilde h \to 1$ (zero temperature), the chemical potential $\mu$ is finite and positive, while $m$ diverges to infinity;
For $\tilde h \to 2$ (infinite temperature) $\mu \to -\infty$, while $m$ is finite and negative.

\begin{figure}[ht!]
\begin{center}
\includegraphics[width=0.7\textwidth,clip]{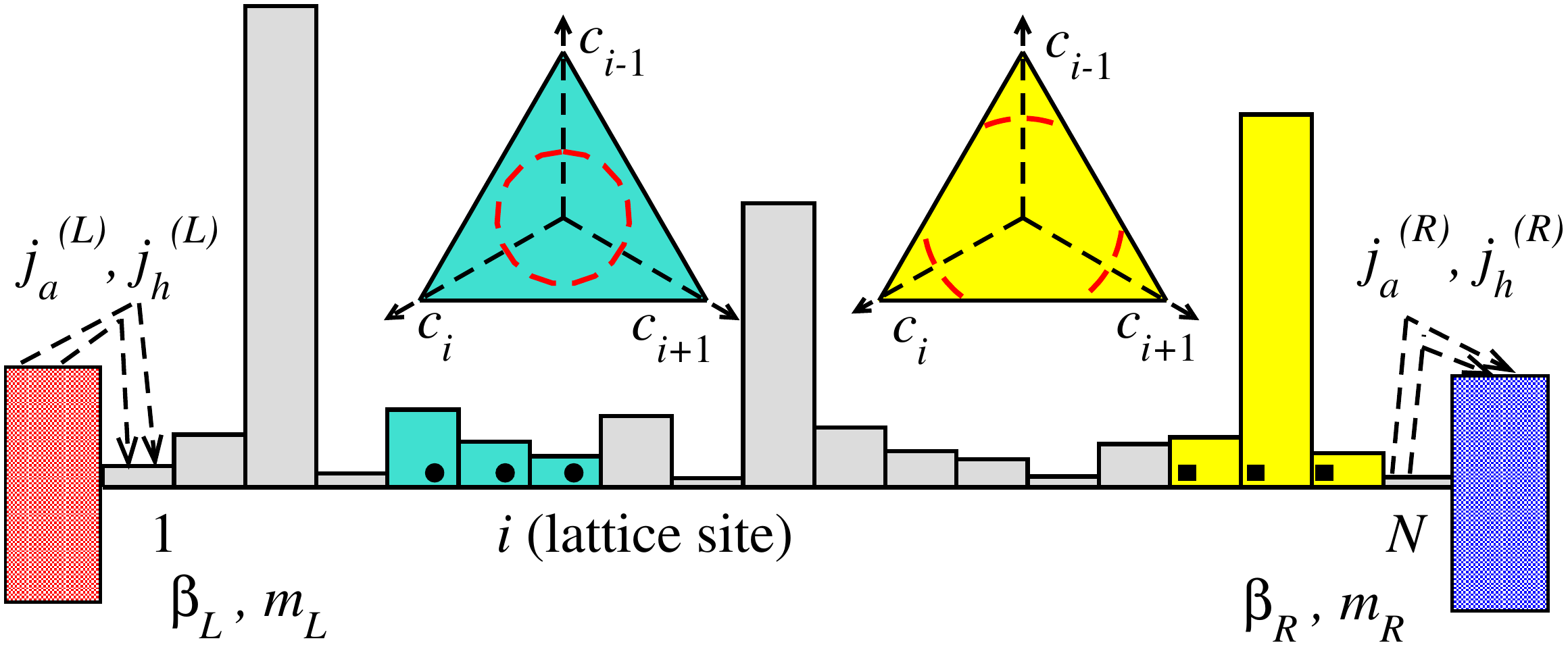}
\caption{The transport setup: a C2C chain steadily interacts with two reservoirs at its boundaries.  
Boundary thermal conditions are specified by the couple of values $\beta$ and $m\equiv \beta\mu$.
$j_a$ and $j_h$ represent the mass and the energy current, respectively.
A NESS is characterized by equal currents at the boundaries, $j_{a,h}^L = j_{a,h}^R$.
Bulk dynamics is implemented using the MMC algorithm (see text)
on random triplets of consecutive sites. Local conservation of energy and
mass restricts the available states in the space $(c_{i-1}, c_i , c_{i+1})$
on the intersection between a plane (colored triangles) and a sphere (not shown),
restricted to the positive octant $c_i\geq 0$. The resulting set of states is either a full circle
(left triangle) or a union of three distinct arcs (right triangle): in both cases the intersection
is given by the red dashed lines. 
%
}
\label{fig:setup}
\end{center}
\end{figure}

The study of out-of-equilibrium properties requires the definition of some dynamical rule.
The C2C dynamics is actually simulated by a Monte Carlo Microcanonical (MMC) algorithm, where
the two conservation laws are implemented for randomly selected triplets of neighbouring sites%
$(i-1,i,i+1)$, so as to satisfy detailed balance~\cite{JSP_DNLS}%
\footnote{Three is the minimal number of sites allowing to satisfy conservation laws and letting
the system evolve. When simulating the system at equilibrium the three sites may not be neighbours,
which speeds up the relaxation to equilibrium~\cite{GIP21}, but in an out-of-equilibrium setup an update rule
among distant sites would generate unphysical couplings between such sites.}. 
In practice this amounts to pass from 
$(c_{i-1},c_i,c_{i+1})$ to $(c'_{i-1},c'_i,c'_{i+1})$, in such a way that (i)
the sum of the three masses and of their squares is conserved and (ii)
that the probability of the transition $\{ c\} \to \{ c'\}$ is equal to the probability of the inverse transition,
$\{ c'\} \to \{ c\}$. 
Geometrically, legal configurations of triplet states lie within the intersection between a plane and a sphere 
in a three-dimensional space, which 
are representative of mass- and energy conservation, respectively.
Since these two constraints define a circle in a three dimensional
space, detailed balance is ensured by picking a random angle.
Depending on the further constraint posed by the mass positivity, $c'_i\geq 0$, 
physically accessible states consist of either a full circle, or the union of three 
disconnected arcs (see Fig.~\ref{fig:setup}).
%
%
%
Here, we have adopted the rule that the new state
should be restricted to the same starting arc, but in~\ref{sec:nopin} we also briefly
consider the case when such restriction is removed.

Out-of-equilibrium (grand canonical) simulations can be 
performed by attaching the two lattice ends to thermal reservoirs.
Customarily, a Monte Carlo grand canonical heat baths~\cite{gotti22} are used.
In this paper, we directly impose the exact equilibrium distributions of masses, which
allows sampling more efficiently the NESS, especially in proximity of the critical line.
In practice we 
extract at random an integer $k \in [1,N]$.
If $k\ne 1,N$ we update the triplet $(k-1,k,k+1)$ as explained here above.
If $k=1$ ($k=N$), we extract a random mass $c\ge 0$ according to the grand canonical
distribution $P(c) \sim \exp{(-\beta^{(i)} c^2+m^{(i)} c)}$
and we assign it to the chosen site. The parameters $\beta^{(i)},m^{(i)}$ define the heat baths attached
to the chain ends ($i=L$ for $k=1$ and $i=R$ for $k=N$), see Fig.~\ref{fig:setup}.
As usual in Monte Carlo simulations, time is measured in units of Monte Carlo moves divided by the system size $N$.

Suitable definitions of mass and energy fluxes can be employed to measure the rate of exchange of
these two quantities from the reservoirs to the chain.
For the left boundary, we define
\begin{eqnarray}\label{eq:flux}
j_a^{(L)}&=& \frac{1}{\tau} \sum_{t_k<\tau} \delta c_1(t_k)\nonumber\\
j_h^{(L)}&=& \frac{1}{\tau} \sum_{t_k<\tau} \delta c_1^2(t_k) \,,
\end{eqnarray}
where $\delta c_1(t_k)$ and $\delta c_1^2(t_k)$ represent respectively the variations  of mass and energy
on the first lattice site produced by reservoir updates occurring at times $0\le t_k\le\tau$. An average over a sufficiently long time window 
$\tau\gg 1$ must be considered.
The definitions of $-j_a^{(R)}$ and $-j_h^{(R)}$ on the right boundary are readily obtained by replacing 
$c_1 \rightarrow c_N$ 
in Eqs.~(\ref{eq:flux}). According to these definitions, fluxes are positive when they flow from left to right.

When a NESS is attained, the conditions $j_a^{(L)}=j_a^{(R)}\equiv j_a$ and $j_h^{(L)}=j_h^{(R)}\equiv j_h$ hold and 
stationary spatial profiles of mass and energy are defined respectively as $a_i = \langle c_i(t)\rangle$
and $h_i= \langle c_i^2(t)\rangle$, where the symbol $\langle\cdot\rangle$ refers to an average over the NESS
distribution.  
 Once plotted parametrically in
the plane $(a,h)$ or in the plane $(m,\beta)$ (through the mapping in Eqs.~(\ref{eq:orig1}-\ref{eq:orig2})) 
the above paths identify a
 ``trajectory" connecting the boundary conditions imposed by the reservoirs.

\section{Onsager coefficients}
\label{sec:Onsager}

The presence of two conservation laws in the C2C model implies the existence of  mass- and energy currents,
which are time and site-independent once the system has reached a NESS.
In the limit of small thermodynamic forces,
local equilibrium sets in and a linear response approach can be adopted.  
Using the standard formulation in terms of the variables $(m,\beta)$, we can write
\begin{eqnarray}
\label{eq.Ons1}
j_a &=& -L_{aa} m_y + L_{ah}\beta_y \\
j_h &=& -L_{ha} m_y + L_{hh}\beta_y, 
\label{eq.Ons2}
\end{eqnarray}
where $L_{uv}$ with $u,v=\{a,h\}$ identify the coefficients of the Onsager matrix~\cite{livi17} and the subscript $y$ denotes a derivative
with respect to the continuous spatial variable $y=i/N$. 

A first important point worth discussing is that the form of the equilibrium equations Eqs.~(\ref{eq:orig1}-\ref{eq:orig2}) implies a scaling form
for the Onsager coefficients.
In fact, if $m\to cm$, then $\beta\to c^2 \beta$ in order to keep constant $z$.
Under the same transformation, see Eqs.~(\ref{eq:orig1}-\ref{eq:orig2}), $(a,j_a) \to (a,j_a)/c$ and $(h,j_h) \to (h,j_h)/c^2$.
Imposing that Eqs.~(\ref{eq.Ons1}-\ref{eq.Ons2}) must be invariant under such scale
transformation, each term on the RHS of Eq.~(\ref{eq.Ons1}) must rescale as $1/c$
and each term on the RHS of Eq.~(\ref{eq.Ons2}) must rescale as $1/c^2$.
Using the scaling assumption 
\be
L_{uv}(m,\beta) = |m|^{\gamma_{uv}} \overline{L}_{uv}(z) 
\label{eq.scalL}
\ee
we obtain the following relations for the scaling exponents $\gamma_{uv}$,
\bea
1 + \gamma_{aa} = -1 &\qquad& 2 +\gamma_{ah} = -1 \nonumber\\
1 + \gamma_{ha} = -2 &\qquad& 2 +\gamma_{hh} = -2 ,
\eea
so that
\bea
\label{eq:scalL2}
L_{aa}(m,\beta) &=& |m|^{-2} \overline{L}_{aa}(z) \nonumber\\
L_{ah}(m,\beta) &=& |m|^{-3} \overline{L}_{ah}(z) \nonumber\\
L_{ha}(m,\beta) &=& |m|^{-3} \overline{L}_{ha}(z) \nonumber\\
L_{hh}(m,\beta) &=& |m|^{-4} \overline{L}_{hh}(z).
\eea

According to Eq.~(\ref{eq.scalL}), the dependence of the Onsager matrix
on the thermodynamic parameters is  determined by
the one-parameter functions $\overline L_{uv}(z)$, with the additional condition 
$\overline{L}_{ah}(z)= \overline{L}_{ha}(z)$
due to the well known symmetry property of the Onsager coefficients~\cite{benenti17rev,livi17}.
By recalling that $m=F_1(z)/a$ and that $z$ is a function of $\tilde h$,
Eqs.~(\ref{eq.scalL}) can be equivalently written in terms of the microcanonical variables $(a,h)$, 
\be
\label{eq:scalL3}
L_{uv}(a,h) = a^{-\gamma_{uv}} L^*_{uv}(\tilde{h}) \,,
\ee
perhaps preferable, as $a$ is by definition positive.

\begin{figure}[ht!]
\begin{center}
\includegraphics[width=0.7\textwidth,clip]{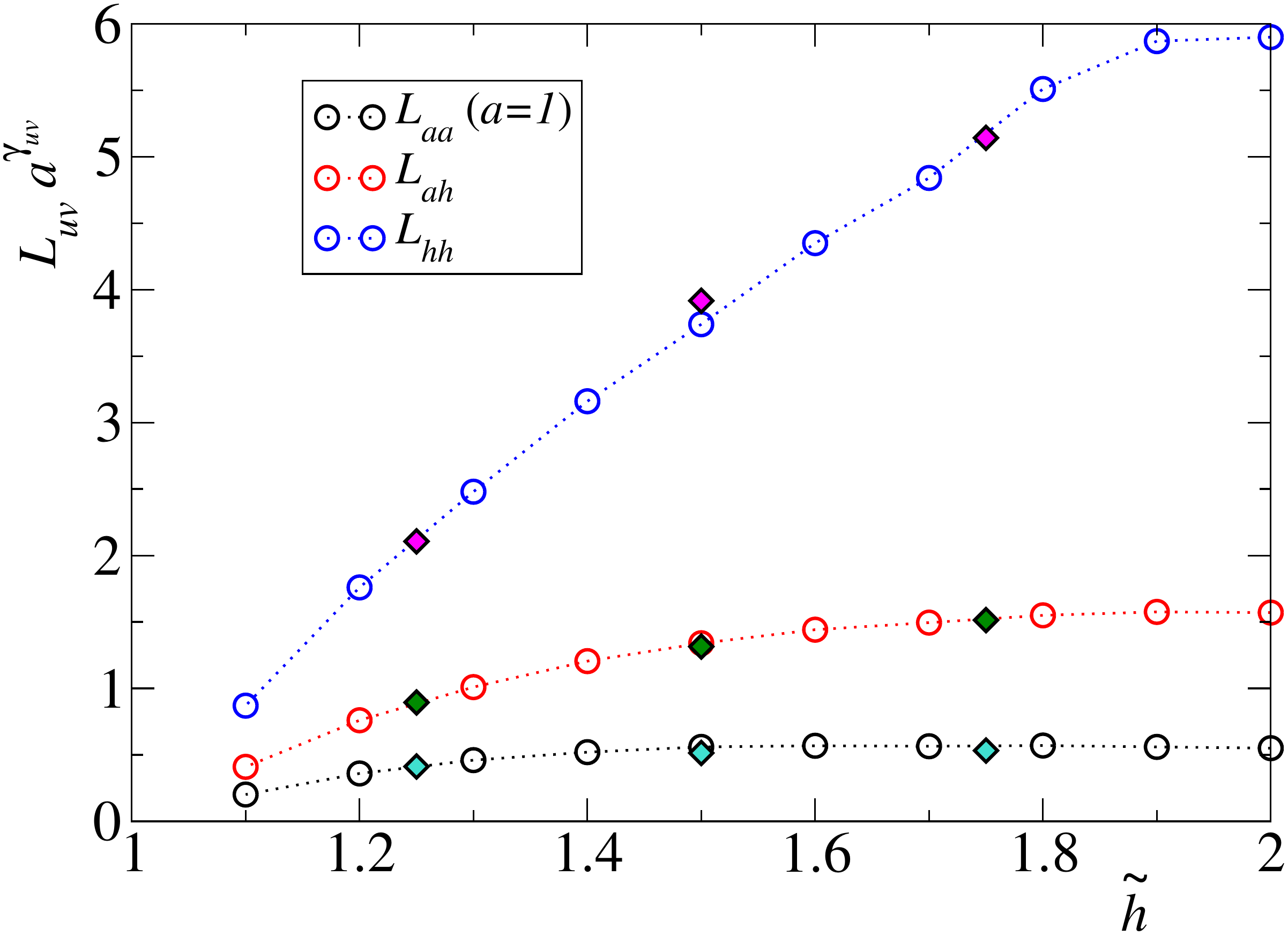}
\caption{
Plots of the Onsager coefficients $L_{aa}$ (lower data), $L_{ah}$ (middle data), and $L_{hh}$ (upper data)
as functions of $\tilde h$.
Circles have been obtained for $a=1$, diamonds for $a=2$.
Dotted lines are guides for the eye.}
\label{fig:scaling}
\end{center}
\end{figure}

In order to test the above scaling, we have plotted $L_{uv} a^{\gamma_{uv}}$ as a function of $\tilde{h}$ for different values of
$a$, see Fig.~\ref{fig:scaling}. 
 Onsager coefficients were computed using Eqs.~(\ref{eq.Ons1}-\ref{eq.Ons2}) for given
 values of the thermodynamic forces and inserting the corresponding values of stationary fluxes determined
 numerically. Since there are four Onsager coefficients (three independent ones) it is necessary to 
 analyze at least two independent paths passing through the same reference point $(m,\beta)$. 
The agreement between circles $(a=1)$ and diamonds $(a=2)$ confirms 
the validity of Eq.~(\ref{eq:scalL3}).
The curves plotted in Fig.~\ref{fig:scaling} 
depend smoothly on $\tilde h$ and there is no divergence for $\tilde h\to 2$.
The behavior of Onsager coefficients in proximity of the critical curve
will be treated in more detail in Sec.~\ref{sec:high_T}.

\subsection{Low temperature limit}
\label{sec:low_T}

The stochastic move of our MMC algorithm does not allow for a straightforward interpretation
of the mass changes.
This is because the three-arcs solution, see Fig.~\ref{fig:setup} and below Eq.~(\ref{eq:zan}),
depends in a complicated way on the initial triplet. 
This is no longer the case if the solution is a full circle:
in this case, any possible final triplet can be obtained by a random rotation
$\theta$, with $0\le \theta < 2\pi$~\cite{JSP_DNLS}.
Hence, the low$-T$, i.e. $\tilde h\to 1$, limit
can be treated analytically. In fact, the ground state of the system
corresponds to equal masses, $c_i \equiv a$, and low$-T$ configurations
are characterized by weak spatial fluctuations of the local mass.
Therefore, in this limit
the intersection between the plane of constant mass and the sphere of constant energy is
a full circle.
As we are going to argue, this allows to obtain some analytical results.

In formulae, the update $(c_1,c_2,c_3) \rightarrow (c'_1,c'_2,c'_3)$ reads
\be
(c'_1,c'_2,c'_3) =R\,(c_1,c_2,c_3)
\ee 
where $R$ is the rotation matrix around the $(1,1,1)$ direction
\be
R=
\begin{pmatrix}
d& f& s\\
s& d& f\\
f& s& d
\end{pmatrix} ,
\ee
with
\begin{eqnarray}
\label{eq:rot}
d&=& \frac{1}{3} +\frac{2}{3}\cos\theta\nonumber\\
f&=& \frac{1}{3}(1-\cos\theta) -\frac{1}{\sqrt{3}}\sin\theta\nonumber\\
s&=& \frac{1}{3}(1-\cos\theta) +\frac{1}{\sqrt{3}}\sin\theta .
\end{eqnarray}

We now write the mass and energy fluxes as~%
\footnote{The factor 2 comes from the fact that
given any pair of consecutive sites, there are two different
triplets which contribute to the flux.}
\be
\label{eq:jaLT}
j_a = 2\langle \overline{c'_1}-c_1 \rangle \quad j_h= 2\langle \overline{\e'_1} - \e_1 \rangle
\ee 
where the symbol $\overline{\cdots}$ refers to the average over the distribution 
of $\theta$ for a given initial state $(c_1,c_2,c_3)$,
while  $\langle \cdots\rangle$ refers to the average over the distribution of
the initial triplet, $(c_1,c_2,c_3)$.
In the general case, not all $\theta$ angles are allowed because of the constraint $c_i '\ge 0$.
However, if the variance of the three masses is not large, then all $\theta$ are allowed
and the average over $\theta$ is trivial~\footnote{More precisely this occurs 
if $\sigma^2 \le \bar c^2/2$, where $\bar c$ and $\sigma^2$
are the mean and the variance of the three initial masses.}.

Let us first consider the expression of the mass flux $j_a$.
The average of the first of Eq.~(\ref{eq:jaLT}) over the distribution of $\theta$ is easily computed and
gives
\bea
\label{eq:jaLT2}
j_a&=
& 2\left(\frac{1}{3} \langle c_1+c_2+c_3\rangle - \langle c_1 \rangle\right) \nonumber\\
&=& 
\frac{2}{3} \langle -2c_1+c_2+c_3\rangle .
\eea
We now suppose that a mass gradient is present in the triplet:
\begin{eqnarray}
\langle c_1 \rangle &=& \langle c_2 \rangle +\Delta_m \equiv a +\Delta_m \nonumber\\
\langle c_3 \rangle &=& \langle c_2 \rangle -\Delta_m \equiv a-\Delta_m .
\end{eqnarray}
Inserting these expressions in Eq.~(\ref{eq:jaLT2}) we finally obtain
\be
j_a =-2\Delta_m .
\ee

Analogous calculations can be performed for the energy flux $j_h$. 
The first average over $\theta$ reads 
\bea
\overline{\e'_1} -\e_1 &=& \overline{[D(\theta)c_1 +M(\theta)c_2+P(\theta)c_3]^2}-\e_1 \nonumber \\
&=& \frac{1}{3}(-2\e_1+\e_2+\e_3) .
\eea
Then, by assuming an energy gradient
\begin{eqnarray}
\langle \e_1 \rangle &=& \langle \e_2 \rangle +\Delta_\epsilon \equiv h +\Delta_\epsilon\nonumber\\
\langle \e_3 \rangle &=& \langle \e_2 \rangle -\Delta_\epsilon \equiv h-\Delta_\epsilon\,,
\end{eqnarray}
we obtain
\be
j_h =-2\Delta_\epsilon .
\ee

Remarkably, in the low temperature limit 
both fluxes do not depend on spatial correlations between sites. 
In a continuum representation we can summarize the above result by writing
\begin{eqnarray}\label{eq:ons_ah}
j_a &=& C_{aa} \,\partial_y a + C_{ah}\,\partial_y h\\
j_h &=& C_{ah} \,\partial_y a + C_{hh}\,\partial_y h\nonumber
\end{eqnarray}
with
\begin{eqnarray}\label{eq:ons_ah2}
C_{aa} &=& -2\\
C_{ah} &=& C_{ha} = 0 \nonumber\\
C_{hh} &=& -2 . \nonumber
\end{eqnarray}
Accordingly, in this representation the two currents are decoupled.

In order to determine the proper Onsager coefficients, we need to map Eq.~(\ref{eq:ons_ah}) onto Eq.~(\ref{eq.Ons1}). 
In practice, it is necessary to express the derivatives $\partial_y a$ and $\partial_y h$ in terms of the thermodynamic
forces $\partial_y \beta$ and $\partial_y m$. In formulae,
\begin{eqnarray}
\label{eq:ons_ah2bis}
\nonumber
j_a &=& (C_{aa}\partial_m a + C_{ah}\partial_m h)\partial_x m + (C_{aa} \partial_\beta a +C_{ah} \partial_\beta h) \partial_x \beta\\
j_h &=& (C_{ah}\partial_m a + C_{hh}\partial_m h)\partial_x m + (C_{ah} \partial_\beta a + C_{hh}\partial_\beta h) \partial_x \beta \nonumber
\end{eqnarray}

By using the relations in Eq.~(\ref{eq:ons_ah2}), Eq.~(\ref{eq:ons_ah2bis}) simplifies to 
\begin{eqnarray}
\nonumber
j_a &=& -2\partial_m a\,\partial_x m -2\partial_\beta a\, \partial_x \beta\\
j_h &=& -2\partial_m h\,\partial_x m - 2\partial_\beta h \partial_x \beta \nonumber
\end{eqnarray}
and we obtain
\begin{eqnarray}
\label{eq:ons_ah3}
L_{aa}=2\partial_m a \quad && L_{ah}=-2\partial_\beta a\\
L_{ha}=2\partial_m h \quad && L_{hh}=-2\partial_\beta h\nonumber .
\end{eqnarray}

The derivatives in Eq.~(\ref{eq:ons_ah3}) are completely determined by the ``equation of state'' of the model, Eq.~(\ref{eq:orig1},\ref{eq:orig2}).
Moreover, the reciprocity property $L_{ah}=L_{ha}$ is recovered by recalling the standard grand canonical relations
 $a=\partial_m \log Z$ and $h=-\partial_\beta \log Z$, where $Z$ is the 
partition function of the model, see Eq.~(\ref{eq:Z}). 
Indeed, given the regularity of $\log Z$, the equality of off-diagonal coefficients follows from 
$\partial_\beta \partial_m \log Z = \partial_m \partial_\beta \log Z$.

\begin{figure}[ht!]
\begin{center}
\includegraphics[width=0.7\textwidth,clip]{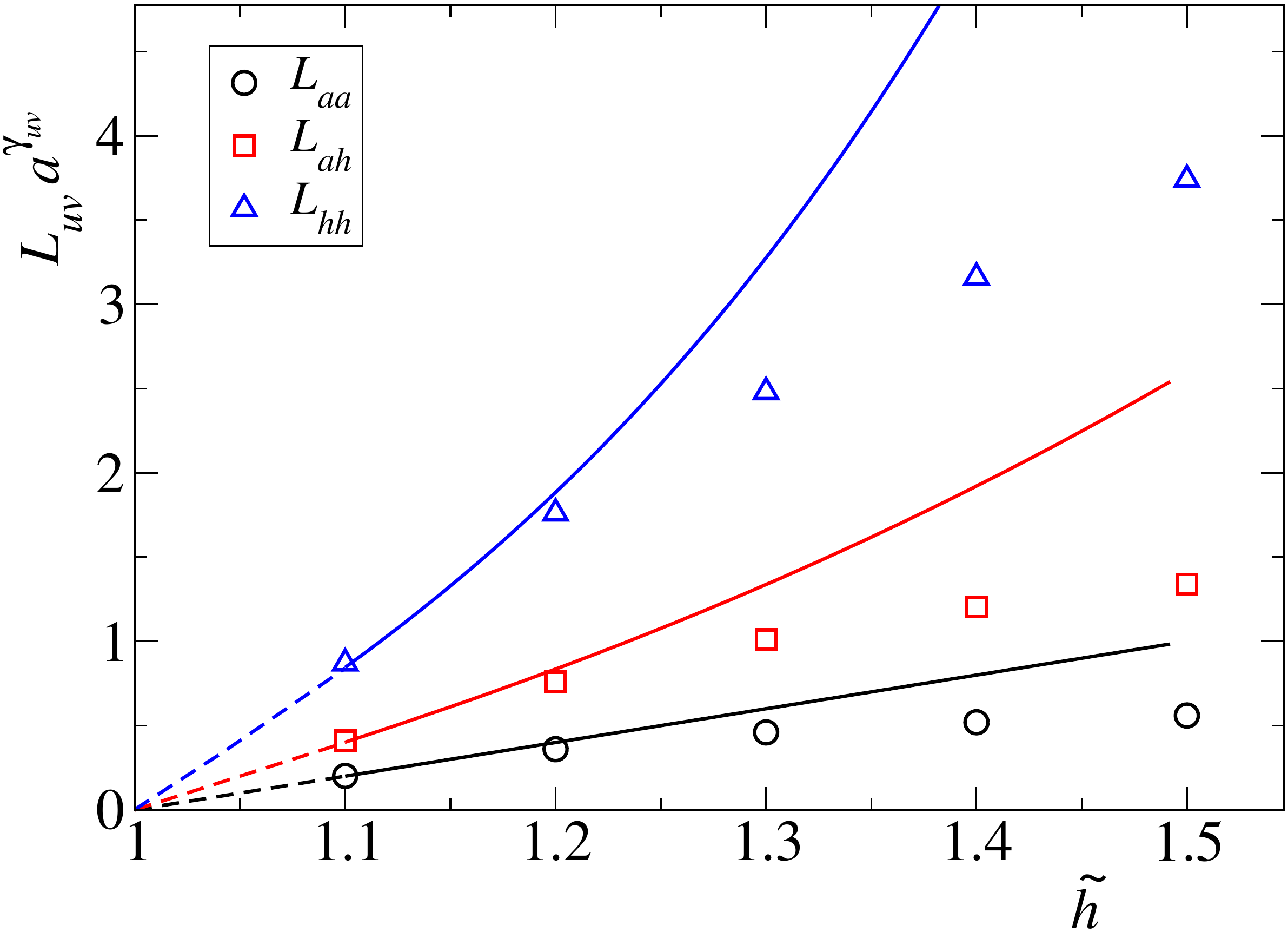}
\caption{
Onsager coefficients versus $\tilde h$. Symbols refer to numerical simulations while solid lines are the analytic
estimates obtained from Eq.~(\ref{eq:ons_ah3}). Lower, middle and upper data refer respectively to 
$L_{aa}$, $L_{ah}$ and  $L_{hh}.$ Dashed lines are obtained from the low-temperature expressions in Eq.~(\ref{eq:ons_ah4}).
Simulations were obtained for $a=1$.
}
\label{fig:l}
\end{center}
\end{figure}

In Fig.~\ref{fig:l} we
compare the Onsager coefficients determined numerically in Fig.~\ref{fig:scaling} (symbols)
with the analytical estimates of Eq.~(\ref{eq:ons_ah3}) (solid lines).  
In the limit of vanishing temperature, $\tilde h\to 1$, it is possible to derive simple expressions of the coefficients by neglecting
the exponential terms in Eq.~(\ref{eq:orig1}). We find
\begin{eqnarray}
\label{eq:ons_ah4}
L_{aa}=2a^2\left(\tilde h-1\right) \quad && L_{ah}=4a^3\left(\tilde h-1\right)\\
L_{ha}=L_{ah} \quad && L_{hh}=4a^4\left(\tilde{h}^2-1\right)\nonumber \,,
\end{eqnarray}
which show that all $L_{uv}$ vanish linearly as $(\tilde h -1)$ in this limit.

Figure~\ref{fig:l} also shows the limits of our analytic approximation of the Onsager coefficients, 
based on the hypothesis that the new state of a triplet can be always found on the full circle:
this hypothesis starts to fail when $\tilde h \gtrsim 1.2$.

It is now interesting to discuss the origin of this discrepancy,
because the passage from a full circle to three arcs has two effects on our MMC algorithm.
When the three masses of a triplet $(c_1,c_2,c_3)$ are sufficiently heterogeneous, the intersection between the plane
of constant mass and the sphere of constant energy is the union of three disjoint arcs rather than
a single connected circle~\cite{JSP_DNLS}. For this class of moves, the analytic result of Eq.~(\ref{eq:ons_ah3})  overestimates the stationary
flux, as it assumes that the rotation angle $\theta$ in Eq.~(\ref{eq:rot}) always varies in $[0,2\pi)$. 
In~\ref{sec:nopin} we clarify that the main contribution to the observed deviations derives from the pinning property of localized states imposed
by the C2C dynamics, i.e. the fact that a sufficiently large peak cannot jump to neighboring sites.
When pinning is removed (see Ref.~\cite{gotti22} for a discussion of this modification of the model), we find a better agreement with 
the analytic estimate, which extends to  higher $\tilde h$ values.

\subsection{High temperature limit}
\label{sec:high_T}

In this section we determine the Onsager coefficients for a point located in the critical curve at infinite temperature,
$(m_0,\beta_0)=(-1,0)$, corresponding to $a_0=1$ and $\tilde h_0=2$.
Such a \textit{critical} point calls for caution because results obtained at finite $T$ might not be valid
and Onsager coefficients might have some nonanalytic behavior.

\begin{figure}
\begin{center}
\includegraphics[width=0.7\textwidth,clip]{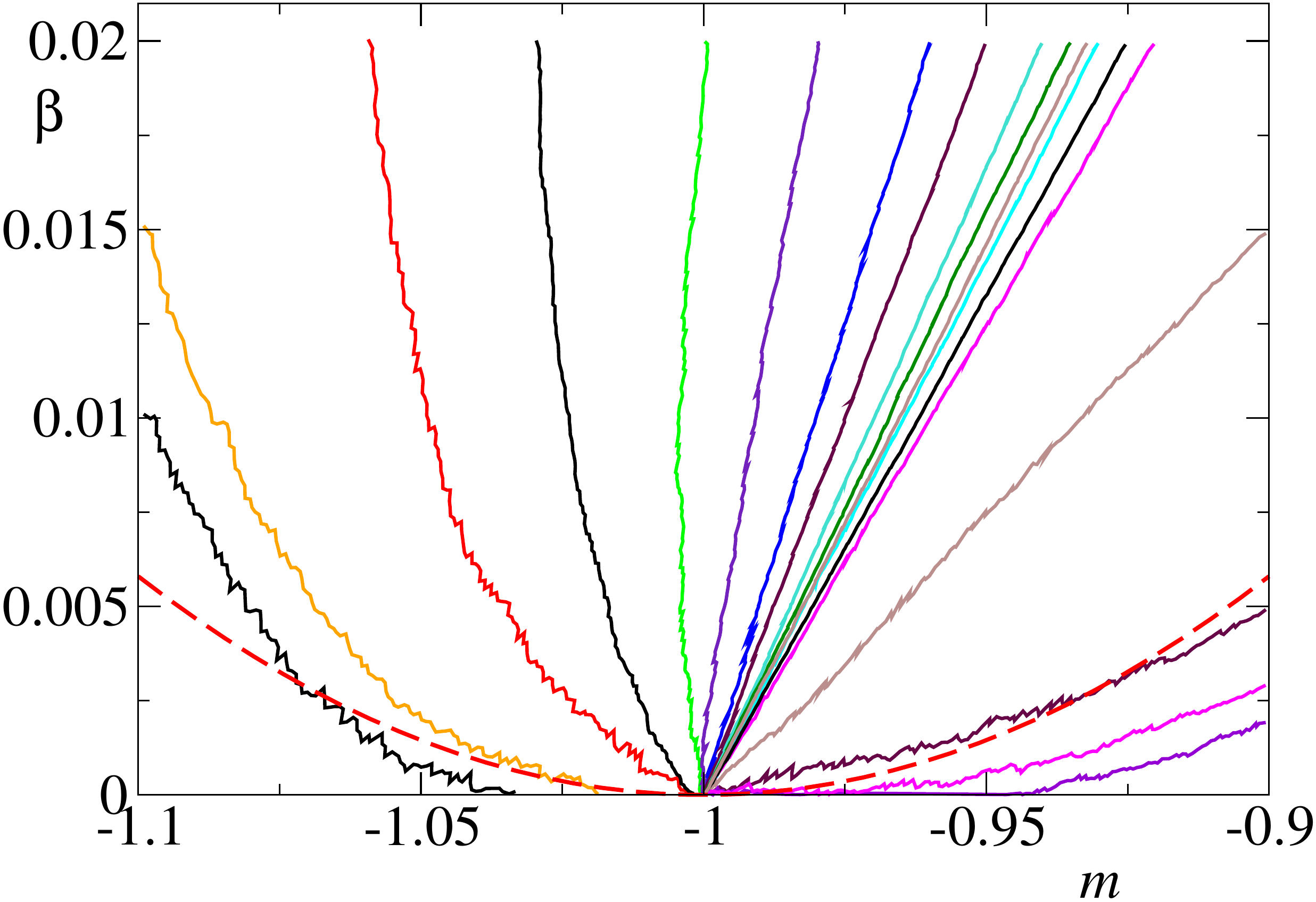}
\caption{Different stationary paths in the $m$, $\beta$ plane. All curves terminate in the infinite-temperature
point $(m_0,\beta_0)$. 
The red dashed line represents the analytic approximation of the path tangent to $\beta=0$ in $m=-1$, see Sec.~\ref{sec:tn}.
 }
\label{fig:mub}
\end{center}
\end{figure}

For this reason we have performed a detailed numerical investigation in order to ensure significantly accurate simulations.
More precisely, we have generated several parametric curves, all starting in $(m_0,\beta_0)$
and terminating in different points $(m_i,\beta_i)$. The resulting paths are plotted in Fig.~\ref{fig:mub}.
All simulations are done in a system of length $N=160$. A comparison with  length $N=320$ (not shown) confirms that these 
results are asymptotic.
In order to extract $\beta$ and $m$ from the numerical simulations, we have made use of Eqs.~(\ref{eq:orig1},\ref{eq:orig2})
with the help of the perturbative expansion in  Eq.~(\ref{eq:zan}).

Fifteen curves are entirely located in the homogeneous $\beta\geq 0$ region and will be employed to
determine the coefficients $L_{uv}(-1,0)$. 
Note that three curves (the two leftmost ones and the rightmost one) cross the critical line at infinite temperature
thus entering the negative-temperature region of the model. We will further investigate this phenomenon in
Sec.~\ref{sec:tn}.

We proceed by first averaging Eqs.~(\ref{eq.Ons1}-\ref{eq.Ons2}) over all simulations, assuming that the
coefficients $L_{u,v}(-1,0)$ do not depend on the slope of the path. The consistency of this assumption will be verified
a-posteriori. We therefore write
\begin{eqnarray}
\langle j_a\rangle &=& -L_{aa} \langle m_y\rangle + L_{ah}\langle \beta_y \rangle\\
\langle j_h\rangle &=& -L_{ha}\langle  m_y\rangle + L_{hh}\langle \beta_y\rangle 
\end{eqnarray}
obtaining
\begin{eqnarray}
\label{eq:Lah}
L_{ah} &=& \frac{\langle j_a\rangle}{\langle\beta_y \rangle}+ \frac{\langle m_y\rangle}{\langle\beta_y \rangle} L_{aa} \\
L_{ha} &=& -\frac{\langle j_h\rangle}{\langle m_y \rangle}+ \frac{\langle \beta_y\rangle}{\langle m_y \rangle} L_{hh} . 
\label{eq:Lha}
\end{eqnarray}
Then we replace these expressions in the original equations (\ref{eq.Ons1}-\ref{eq.Ons2}), 
now indexed by $i=1,\dots,15$ to clarify that each one refers to
a different parameteric curve shown in Fig.~\ref{fig:mub},
\begin{eqnarray}
-j_a^i \langle \beta_y\rangle + \langle j_a\rangle \beta_y^i &=&
L_{aa}\left (m_y^i \langle \beta_y \rangle - \langle m_y\rangle \beta_y^i  \right ) \label{eq:para1}\\
-j_h^i \langle m_y\rangle + \langle j_h\rangle m_y^i &=&
L_{hh}\left (m_y^i \langle \beta_y \rangle - \langle m_y\rangle \beta_y^i  \right ) \label{eq:para2}.
\end{eqnarray}

\begin{figure}
\begin{center}
\includegraphics[width=0.7\textwidth,clip]{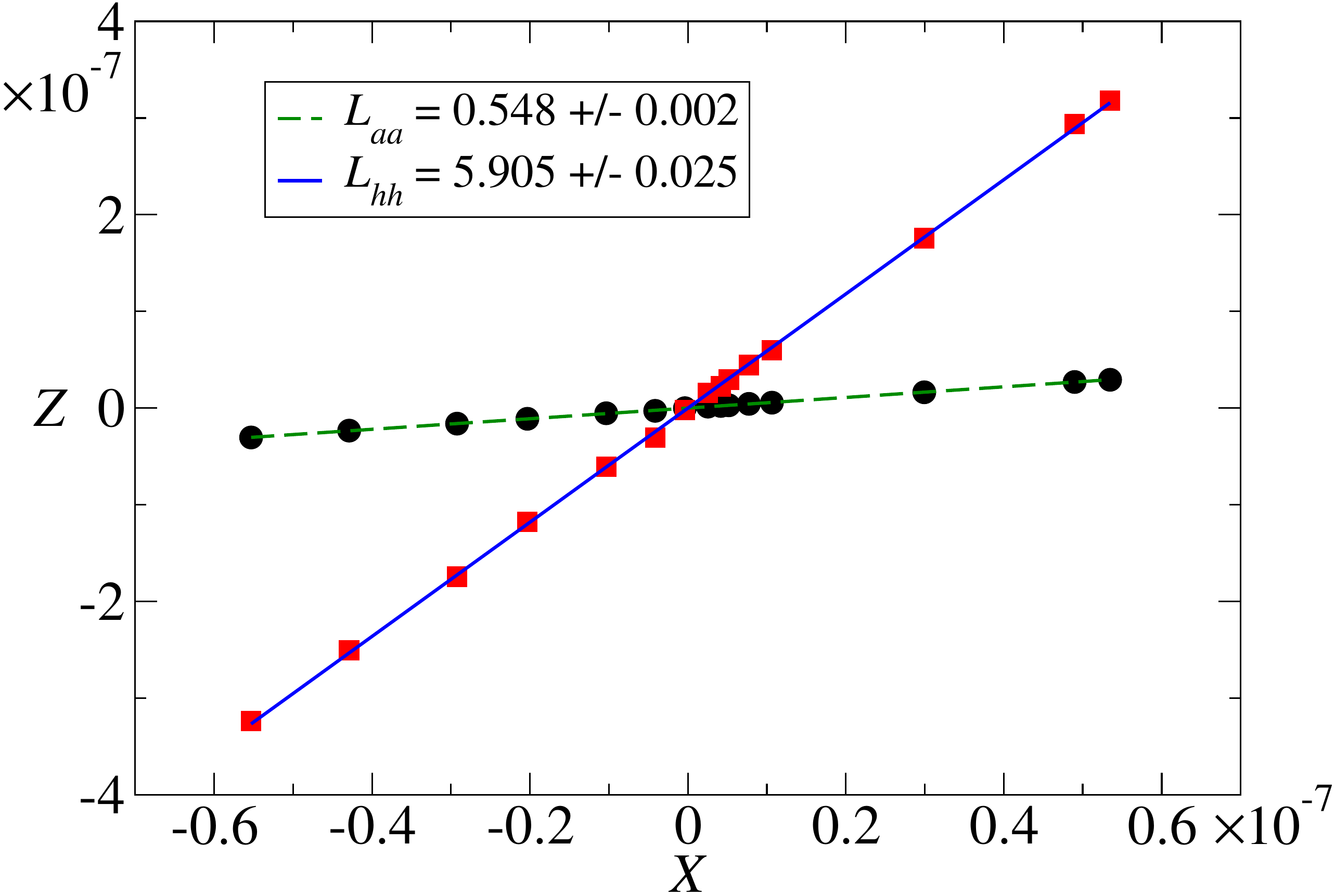}
\caption{Inferring the diagonal elements of the Onsager matrix in the space $(X,Z)$, see Eqs.~(\ref{eq:para1}-\ref{eq:para2}).}
\label{ons1}
\end{center}
\end{figure}

The two equations are of the type $Z^i = L_{uv} X^i$, where $X^i= m_y^i \langle \beta_y \rangle - \langle m_y\rangle \beta_y^i$.
The resulting data are reported in Fig.~\ref{ons1} in the space $(X,Z	)$.
As expected, the data are well aligned along straight lines; their slopes yield the two diagonal coefficients of
the Onsager matrix. By then using the formulas (\ref{eq:Lah}-\ref{eq:Lha}) 
for the off-diagonal elements, we find that
$L_{ah} \sim 1.57\pm 0.03$ and $L_{ha}\sim 1.59\pm0.03$, 
compatible with the theoretical expectation that they must coincide.

By now invoking the scaling form of the Onsager coefficients of Eq.~(\ref{eq:scalL2}), we can extend the above result
to the whole infinite-temperature line. Since the variable $z=m/\sqrt{\beta}$ is constant along this line, and equal to minus
infinity, we obtain
\be
\label{eq:Linf}
L_{uv} (m,\beta=0) = |m|^{\gamma_{uv}} L_{uv}(-1,0),
\ee
where the coefficients $L_{uv}(-1,0)$ have just been determined. We can therefore conclude that the Onsager matrix remains
finite and well-defined on the $\beta=0$ line.

\section{Spatial correlations}
\label{sec:corr}

The analytic approach discussed in Sec.~\ref{sec:low_T} 
clarifies that in the low-temperature regime correlations do not play any role for the determination of
the Onsager coefficients. 
On the other hand, it is reasonable to expect that for sufficiently high temperatures 
transport coefficients do depend on nonequilibrium correlations. 
In this section we analyze the role of correlations and quantify their importance for the coupled transport problem.
For this purpose we focus on a regime close to $\beta=0$ and consider the following setup.
The reservoir on the right boundary imposes $\beta_R=0$ and $m_R=-1$, while
the left reservoir imposes $m_L=m_R$ and $\beta_L=\Delta\beta$, with $\Delta\beta=0.1$
(this corresponds to a line approximately vertical in Fig.~\ref{fig:mub}).
To keep the amplitude of finite-size effects under control, two
lattice lengths $N=50$ and $N=100$ are here compared.

We compute the covariance matrix
\be
C_{ij}=\langle c_i c_{j} \rangle - \langle c_i \rangle  \langle c_j \rangle
\ee  
where $\langle \cdot \rangle$ refers, as before, to the average over the nonequilibrium stationary measure.
The diagonal elements $C_{ii}= h_i  -a_i^2 $ correspond to the local variance of the mass along the chain and do not 
provide information on correlations.

\begin{figure}[ht!]
\begin{center}
\includegraphics[width=0.7\textwidth,clip]{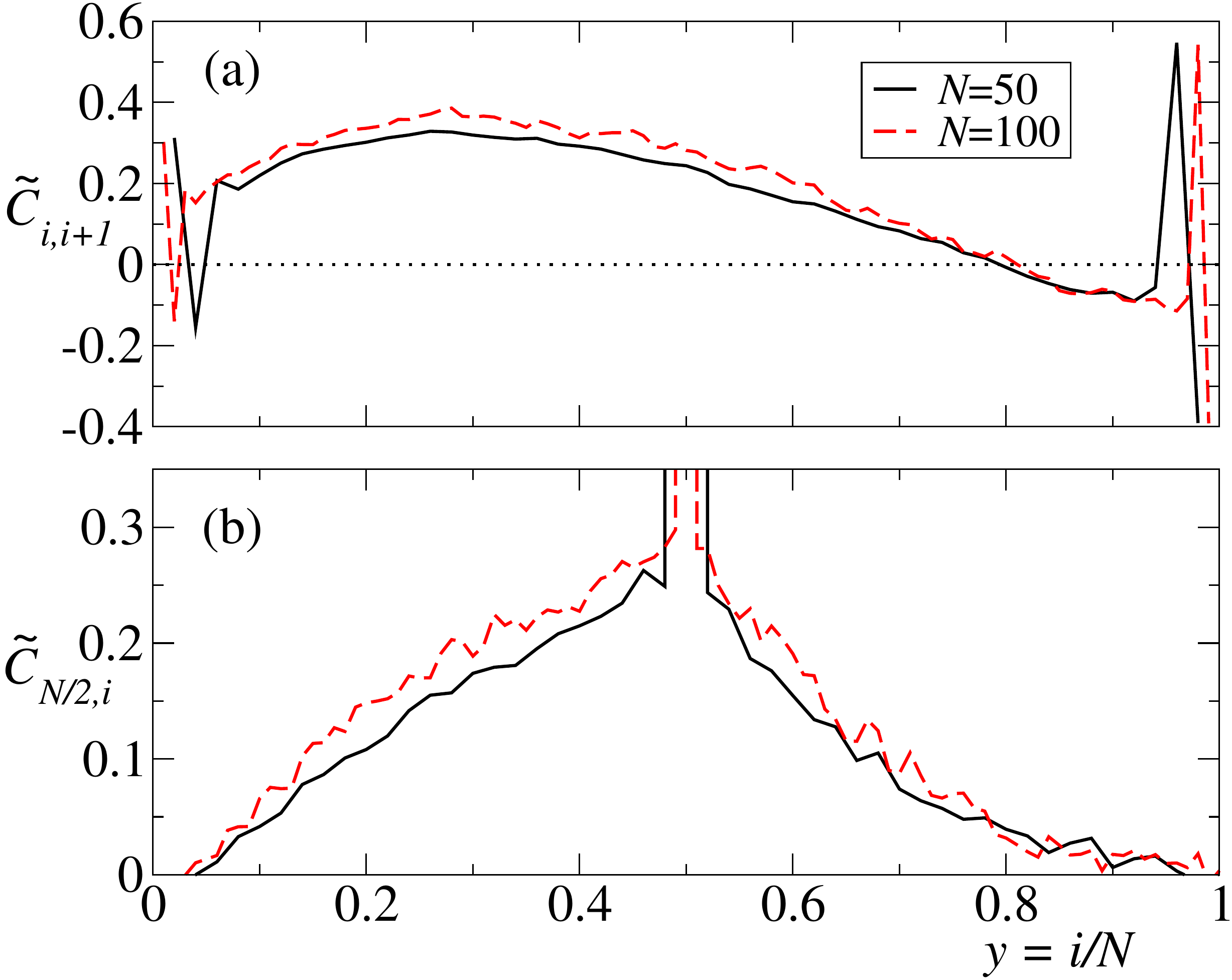}
\caption{(a) Upper diagonal of the rescaled correlation  matrix $\tilde C_{ij}$ for $N=50$ (solid line) and $N=100$ (dashed line).
The dotted line highlights the zero baseline.
(b) Rescaled correlation matrix $\tilde{C}_{N/2,i}$ for the same lattice sizes.
}
\label{fig:corr}
\end{center}
\end{figure}

Off-diagonal elements of $C_{ij}$ are expected to vanish as the gradient of $\beta$ goes to zero. For this reason, it is convenient
to rescale the correlation matrix with the gradient of $\beta$,
\be
\tilde{C}_{ij} =  N C_{ij}/\Delta\beta . 
\ee
In Fig.~\ref{fig:corr} we show the main features of $\tilde{C}_{ij}$, as found from  numerical simulations. 
In panel (a) we report the nearest-neighbor correlations (located in the upper diagonal $\tilde{C}_{i,i+1}$)
as a function of the rescaled position $y=i/N$.
A nontrivial correlation pattern is obtained, characterized by an asymmetric distribution of positive and negative correlations. 
In panel (b) we show the behavior of $\tilde{C}_{k,i}$ while moving along the entire row corresponding to the central lattice site $k=N/2$. 
Similarly to other nonequilibrium models (see e.g. Ref.~\cite{Bertini2009}), long-range correlations 
are found of amplitude $1/N$ across the entire system.

\begin{figure}[ht!]
\begin{center}
\includegraphics[width=0.7\textwidth,clip]{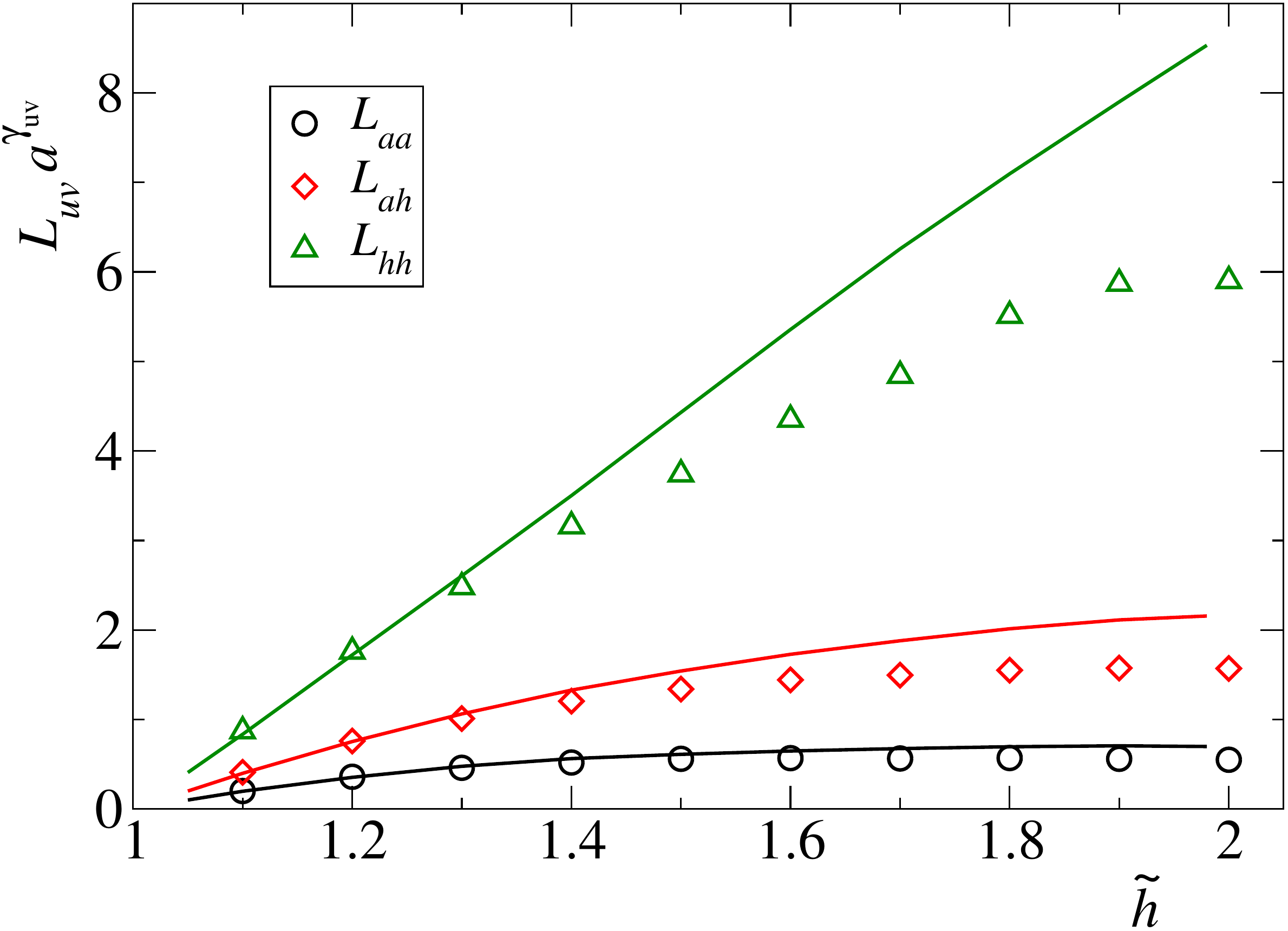}
\caption{Onsager coefficients versus $\tilde h$. Symbols refer to  numerical simulations of the MMC model
(same data of Fig.~\ref{fig:scaling}), while
 solid lines are obtained for the fully uncorrelated dynamics, see text. Lower, middle and upper data series refer
 respectively to $L_{aa}$, $L_{ah}$ and $L_{hh}$.
}
\label{fig:cor_ons}
\end{center}
\end{figure}

To better understand the role of correlations for the Onsager matrix, we have considered a modified MMC dynamics restricted 
to a triplet $(N=3)$
in which correlations are intentionally suppressed~\footnote{In the absence of correlations there is no need to consider larger system sizes.}.
 This can be realized by imposing on each site of the triplet independent distributions
of the local masses, where the distribution on site $i$ is defined by parameters $(\beta_i, m_i)$ chosen in order to produce given constant gradients. 
In Fig.~\ref{fig:cor_ons} we compare the three Onsager coefficients for the full MMC model (symbols) with 
those corresponding to the uncorrelated model (solid lines). As expected, in the low-temperature region correlations are very small for 
the MMC 
dynamics and $L_{u,v}$ are well described by the fully uncorrelated model. On the other hand, for larger $\tilde h$
correlations tend to decrease the values of the Onsager coefficient with respect to the uncorrelated limit.  
This effect is maximal for the infinite-temperature point $\tilde h=2$ and clarifies that the peculiar structure of $L_{uv}$ 
found in this limit depends at least in part on nonequilibrium correlations.

\section{Spontaneous emergence of negative temperatures}
\label{sec:tn}

In Ref.~\cite{gotti22}, it was shown that nonequilibrium stationary paths can enter the
negative-temperature region even when the reservoirs at the chain boundaries impose
positive temperatures. As a result, a new kind of condensation phenomenon may arise, produced
exclusively in nonequilibrium conditions. In this section, we revisit this process from the
point of view of Onsager theory, deriving a perturbative expression of the limiting form of the paths and,
accordingly, the condition for them to enter the negative$-T$ region.
  
We start the analysis by focusing on the shape of the paths that are nearly tangent to the $\beta=0$ line,
see e.g. the second rightmost curve in Fig.~\ref{fig:mub}.
Each stationary path is by definition characterized by constant mass $(j_a)$ and energy $(j_h)$ currents.
It is convenient to take their ratio $\rho$ because we can get rid of the explicit spatial dependence
of $\beta$ and $m$. 
In fact, from Eqs.~(\ref{eq.Ons1}) and (\ref{eq.Ons2}),
\begin{equation}
\label{eq:ratio}
\rho = \frac{ j_a}{ j_h }  = \frac{L_{aa} - L_{ah}\beta_m}{L_{ha}- L_{hh}\beta_m} =
\frac{
|m|\overline{L}_{aa} (z) -
\beta_m \overline{L}_{ah} (z)}
{|m|\overline{L}_{ha}(z)  -
\beta_m\overline{L}_{hh} (z)} |m|
\end{equation}
where $\beta_m = d\beta/dm = \beta_y/m_y$.

A first relevant consequence of the above relation is that the isothermal line $\beta(m)=0$ cannot correspond to a stationary path.
Indeed, along  this line	 Eq.~(\ref{eq:ratio}) would write
\be
\label{eq:ratio2}
\rho = \frac{\overline{L}_{aa} (-\infty)}{\overline{L}_{ha}(-\infty)} |m| \; .
\ee
Since the ratio of Onsager coefficients is finite along the
critical line $\beta=0$ (see Eq.~(\ref{eq:Linf})),
the ratio $\rho$ would grow linearly with $|m|$, contradicting the physical condition of a constant $\rho$ along a NESS path.

We now relax the condition $\beta(m)=0$ and investigate the occurrence of tangent paths.
More precisely we assume
\be
\beta(m) = g(m-m_0)^2 \equiv g \delta^2\,; \quad m= m_0 + \delta\,;\quad \delta\ll 1,
\label{eq:parab}
\ee
where $g$ is a coefficient determining the openness of the parabolic shape.
In the vicinity of $m=m_0$, $|\beta_m|\ll 1$.
Under  this approximation, Eq.~(\ref{eq:ratio}) can be written as
\be
\label{eq:ratio3}
\rho =
\frac{\overline{L}_{aa}(\sqrt{\beta}/m)}{\overline{L}_{ha}(\sqrt{\beta}/m)}|m|  + 
\left (\frac{\overline L_{aa}(\sqrt{\beta}/m)\overline L_{hh}(\sqrt{\beta}/m)}%
{\overline L_{ah}^2(\sqrt{\beta}/m)}-1\right ) \beta_m ,
\ee
where we have expressed the Onsager coefficients $\overline{L}_{uv}$ as functions
of $w\equiv 1/z = \sqrt{\beta}/m$, rather than as functions of $z$, which diverges
in the limit $\beta=0$.

By inserting the parabolic Ansatz for $\beta(m)$ into Eq.~(\ref{eq:ratio3}) and retaining all terms up to order $\delta$, we obtain
\be
\label{eq:ratio4}
\rho = \frac{\overline{L}_{aa}(\sqrt{g}\delta/m_0)}{\overline{L}_{ha}(\sqrt{g}\delta/m_0)} (|m_0| - \delta) + 2g
\left ( \frac{\overline{L}_{aa}(0)\overline{L}_{hh}(0)}{\overline{L}_{ah}^2(0)}-1\right ) \delta ,
\ee
where we have made explicit that the ratio of Onsager coefficients multiplying $(|m_0| -\delta)$ should be
evaluated in $w=(\sqrt{g}/m_0)\delta$. Eq.~(\ref{eq:ratio4}) can be further simplified by considering
the linear expansion
$\overline{L}_{uv}(w)= \overline{L}_{uv}(0) + \overline{L}'_{uv}(0) w$.
The derivative $\overline{L}'_{uv}(0)$ is conveniently determined
passing through the variable $\Delta = 2 - \tilde h$,
\bea
\label{eq:ratio5}
\left. \overline{L}'_{uv}(w)\right|_{w=0} &=&
\left. \overline{L}'_{uv}(\Delta)\right|_{\Delta=0} \left.\frac{d\Delta}{dw}\right|_{w=0}\\ 
&=& \left. \overline{L}'_{uv}(\Delta)\right|_{\Delta=0} 2 w|_{w=0} = 0,
\eea
where the result follows from the combined numerical observation that: (i)
all Onsager coefficients have a finite derivative with respect to $\vec h$ (i.e., with respect to $\Delta$);
(ii) $(d\Delta/dw)_{w=0} = 2w|_{w=0} = 0$. 

As a result, the $\delta-$dependence of the  Onsager coefficients can be neglected to this order and we obtain
\be
\rho = \frac{\overline{L}_{aa}(0)}{\overline{L}_{ha}(0)} (|m_0| - \delta) + 2g
\left ( \frac{\overline{L}_{aa}(0)\overline{L}_{hh}(0)}{\overline{L}_{ah}^2(0)}-1\right ) \delta .
\ee

For this equation to be valid, it is necessary that $\rho$ is independent of $\delta$, therefore
\be
\rho = \frac{\overline{L}_{aa}(0)}{\overline{L}_{ha}(0)} |m_0| 
\ee
and
\be
g =  \frac{1}{2}
\frac{\overline{L}_{ah}(0)\overline{L}_{aa}(0)}{\det \overline{L}(0)} \; .
\label{eq:g}
\ee
The first condition determines the flux ratio along a path crossing tangentially the $\beta=0$ line
in $m=m_0$. By inserting the value of the coefficients determined from the simulations, 
we obtain that $\rho\approx -0.348$
for $m_0=-1$. 
This value is consistent with the ratio observed in eventually tangent paths,
see the second leftmost (orange) curve in Fig.~\ref{fig:mub}, where $\rho\approx-0.344$.  
The second condition determines the concavity of the path. 
Interestingly, it is independent of $m_0$, meaning that the concavity is
constant along the $\beta=0$ line.
More precisely, we find that $g\approx 0.58$, 
in agreement with the concavity of the various paths, see the red dashed line in Fig.~\ref{fig:mub}.

\begin{figure}[ht!]
\begin{center}
\includegraphics[width=0.7\textwidth,clip]{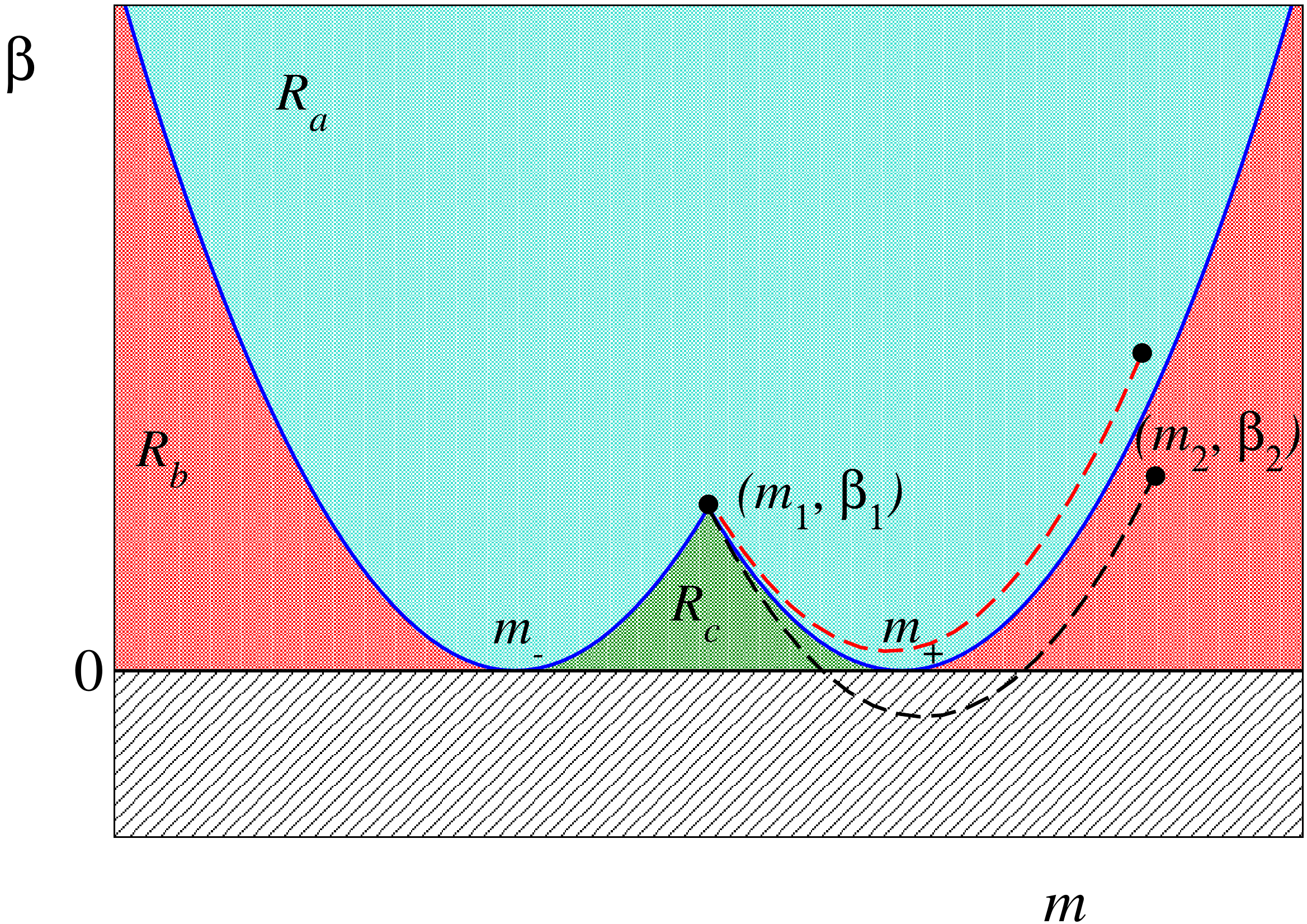}
\caption{
Stationary paths starting from $(m_1,\beta_1)$ can be of three types, depending on the region where they terminate.
If the second reservoir $(m_2,\beta_2)$ is located in $R_a$ ($R_c$), all the path is contained in $R_a$ ($R_c$), and negative
temperatures do not appear (red dashed line).
If the second reservoir is located in $R_b$, the path is initially contained in $R_c$, then it attains the
condensed phase (grey region), and finally it re-enters the positive$-T$ region in $R_b$ (black dashed line). 
$m_\pm$ are the values of $m$ where the limiting paths starting from $(m_1,\beta_1)$ are tangent to the critical line $\beta=0$.
}
\label{fig:negT}
\end{center}
\end{figure}

We now focus our attention on the emergence of paths entering the $T<0$ condensed phase.
Let us suppose that one end of the chain is thermalized at $(m_1,\beta_1)$ (see Fig.~\ref{fig:negT}). 
Two NESS paths depart from this point, that are tangent to the infinite temperature line.
If $\beta_1$ is small, we can rely on the parabolic approximation in Eq.~(\ref{eq:parab}) writing
$\beta_{\pm} = g(m-m_\pm)^2$, where $g$, given by Eq.~(\ref{eq:g}), is the same in both curves.
By imposing that the parabolas pass through $(m_1,\beta_1)$, it follows that
 $m_\pm = m_1 \pm \sqrt{\beta_1/g}$. It is easily seen that these 
two curves partition the parameter space into three regions $R_a$, $R_b$, and $R_c$  (see Fig.~\ref{fig:negT}).
If and only if the other end of the system is located in the region $R_b$, the corresponding path enters the 
negative temperature region; otherwise, the entire path is characterized by positive temperatures. This property follows from
the fact that the family of all stationary paths departing from $(m_1,\beta_1)$ cannot cross the two limiting 
parabolas.
Indeed, upon calling $(m_c,\beta_c)$ the  point of intersection, this would imply the existence of two distinct
paths connecting $(m_1,\beta_1)$ with $(m_c,\beta_c)$. However, ergodicity implies the existence of a single path, the one
 minimizing dissipation~\cite{Onsager}.~\footnote{A path starting in (-1,0) is seen in
Fig.~\ref{fig:mub}(a) to cross the critical parabola, because the parabola is only an approximation valid for vanishing $\beta$.
} 
Formally, if the two reservoirs are located in $(m_1,\beta_1)$ and $(m_2,\beta_2)$, with $m_2 > m_1$,
the path enters the condensed phase if
\be
m_2 > m_1 + \sqrt{\frac{\beta_1}{g}}
\ee
and
\be
\beta_2 < g\left( m_2 -m_1 - \sqrt{\frac{\beta_1}{g}}\right)^2 .
\ee
These conditions are exact in the limit of large temperatures, i.e. for vanishing $\beta_{1,2}$. In  Fig.~\ref{fig:negT} we 
show qualitatively a path entering the condensed region (black dashed line) and a path fully confined in the positive-temperature
region (red-dashed line). 
Similar considerations could be done in the microcanonical parameter space $(a,h)$, using
Eqs.~(\ref{eq:orig1}-\ref{eq:orig2}).
Here we limit to report the expression of the steady path tangent to the critical curve $h=2a^2$
in the point $(a_0,2a_0^2)$:
\be
h(a) = 2 a_0^2 + 4a_0(a-a_0) + (2-4g)(a-a_0)^2 .
\label{eq.ccha}
\ee
Since $g\simeq 0.58$, the coefficient of the quadratic term is negative and the curvature of the
limiting path is therefore opposed to the positive curvature of the critical line.

Non-monotonic temperature profiles are typically
observed in one-dimensional Hamiltonian models in the presence of thermomechanical forces~\cite{iacobucci11,iubini12,Iubini2016_NJP}.
Physically, it is a manifestation of the Joule effect, i.e. the heating of a wire induced by the flow of an
irreversible current of mass (or charge). Here, the effect is extreme, as the inner temperatures become so large as
to become negative.
In more quantitative terms, the local heat production rate $\dot Q$ due to Joule heating can be expressed as (see Eq.~(20)
of Ref.~\cite{benenti17rev})
\be
\dot Q= \frac{j_a^2}{\beta L_{aa} } .
\label{eq:joule}
\ee
In the vicinity of the critical line, still neglecting the dependence of $L_{aa}$ on $\delta$, Eq.~(\ref{eq:joule}) rewrites as
\be
\dot Q= \frac{ j_a^2}{\overline{L}_{aa}(0)} \frac{m^2}{\beta} \,,
\ee
where $j_a$ and $m$ are finite, therefore clarifying that $\dot Q$ diverges with temperature. 
On the other hand, the corresponding contribution to entropy production rate, $\dot Q \beta$,
remains finite.

For the sake of completeness it is worth mentioning that, as shown in Ref.~\cite{gotti22}, 
paths crossing the critical line, may no longer be characterized by a stationary dynamics.
This phenomenon, however, does not affect the path shape in the positive-temperature region.
Strictly stationary paths are, instead obtained, if the 
unrestricted variant of the model described in~\ref{sec:nopin} is adopted~\cite{gotti22}.

\section{Conversion efficiency}
\label{sec:conversion}

Coupled transport can be quantified in terms of the Seebeck coefficient defined as~\cite{benenti17rev} 
\be
\label{eq:seeb}
 S\equiv\beta\frac{L_{ah}}{L_{aa}} -m\,.
\ee
By using the scaling relations for the Onsager coefficients, see Eq.~(\ref{eq:scalL2}), this expression can be rewritten in the form
\be
  S= m \left(\mbox{sign}(z)\frac{\overline{L}_{ah}(z)}{z^2\overline{L}_{aa}(z)} -1  \right)\,,
\ee
therefore, in analogy with $L_{uv}$, it is sufficient to study $S/m$ as a function of $z$, or equivalently $Sa$ as a function of $\tilde h$.

\begin{figure}[ht!]
\begin{center}
\includegraphics[width=0.7\textwidth,clip]{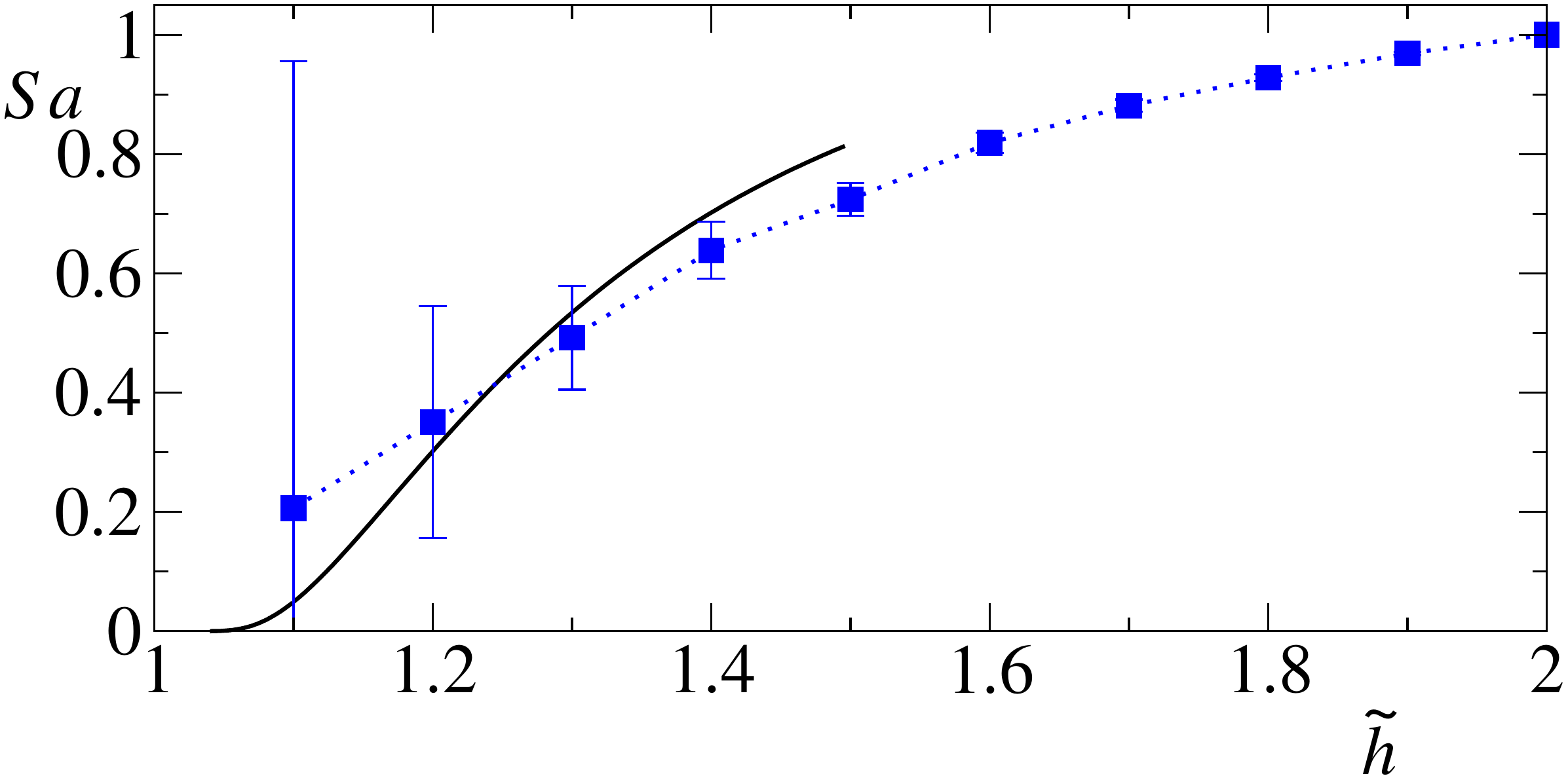}
\caption{
Scaled Seebeck coefficient  $Sa$ versus $\tilde h$. Symbols refer to numerical simulations performed at $a=1$ while
the solid line shows the low-energy analytic prediction obtained from Eq.~(\ref{eq:ons_ah3}). 
Vertical bars are an estimate of numerical errors, see the main text.
}
\label{fig:SD}
\end{center}
\end{figure}

In Fig.~\ref{fig:SD} we show the behavior of $Sa(\tilde h)$ in the whole range $1\leq \tilde h\leq 2$ as obtained from 
numerical simulations (open symbols).  We find that the Seebeck coefficient is positive and monotonically increasing with $\tilde h$. 
We also show as a solid line the analytic estimate obtained from Sec.~\ref{sec:low_T} valid in the low-energy limit.
From this result, we find that $Sa(\tilde h)$ vanishes as $\tilde h\to 1$. 
The deviations observed in the numerical data in this regime  are mostly due to numerical uncertainties in the determination of
 $L_{aa}$ and $L_{ah}$, which are amplified
by the large value of $\beta$ in the definition~(\ref{eq:seeb}), as highlighted by the increasing error bars
with decreasing $\tilde h$.
In the opposite limit $\tilde h\to 2$, $Sa(\tilde h)$ converges to $1$, as immediately found from Eq.~(\ref{eq:seeb}) for
$\beta\to 0$ and $ma\to-1$, see Eq.~(\ref{eq:orig2}).

In the presence of  coupled transport, a measure of the efficiency of conversion of one current into another
is provided by the dimensionless figure of merit
\be
\label{eq:zt}
ZT=	\frac{(L_{ah}-\frac{m}{\beta} L_{aa})^2}{\det{L}} 
\ee
and by the related conversion efficiency
\be
\frac{\eta}{\eta_C}=\frac{\sqrt{ZT+1}-1}{\sqrt{ZT+1}+1}\,,
\ee
where $\eta_C$ is the Carnot efficiency, see~\cite{benenti17rev} for details. 
The ratio $\eta/\eta_C$ increases from $0$ for $ZT\ll 1$ to $1$ for $ZT\gg 1$.

\begin{figure}[ht!]
\begin{center}
\includegraphics[width=0.7\textwidth,clip]{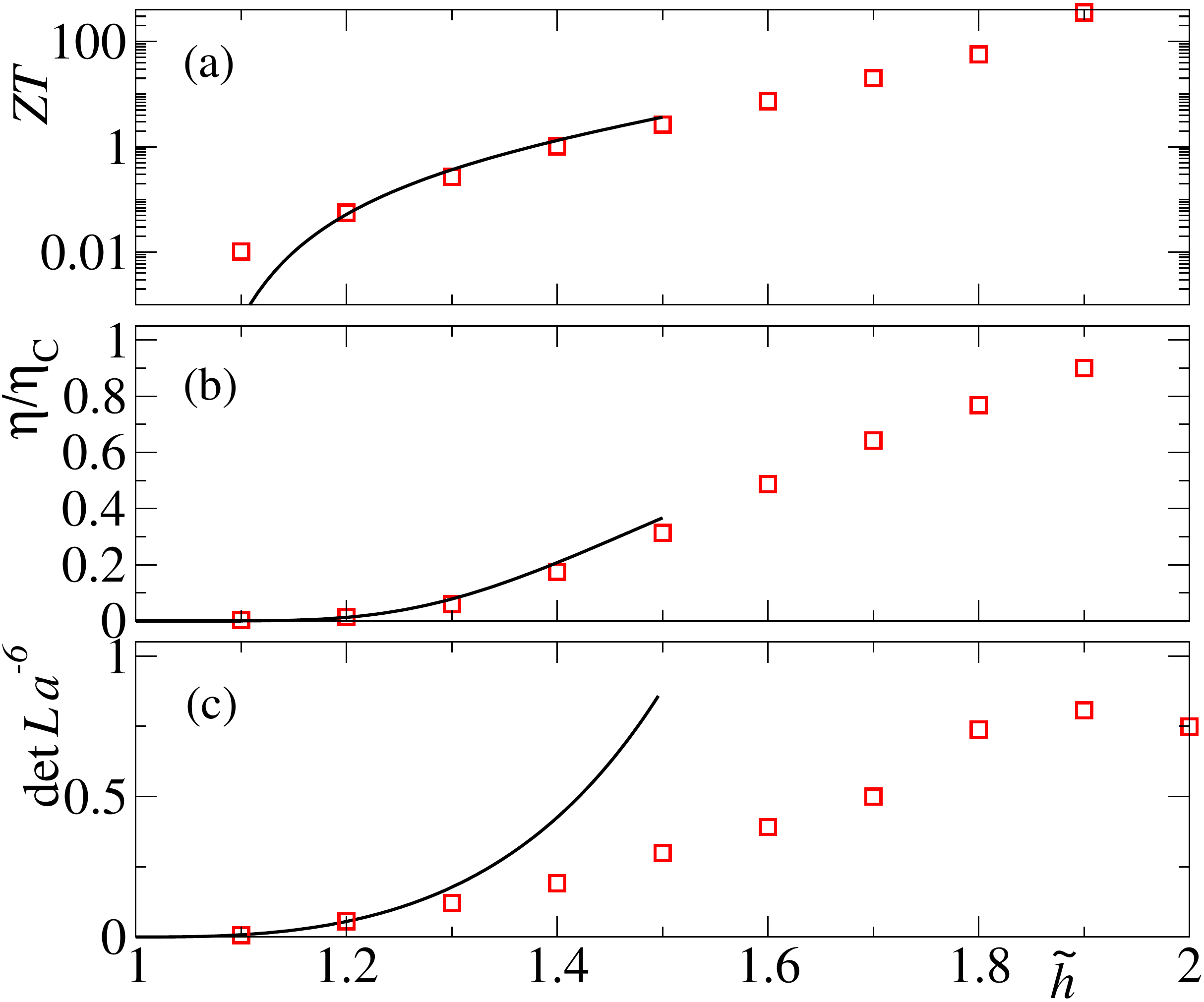}
\caption{
(a)~Figure of merit $ZT$; (b)~Conversion efficiency; (c) Determinant of the Onsager matrix versus $\tilde h$.
 Symbols refer to numerical simulations while solid curves illustrate the analytic prediction obtained from
 Eq.~(\ref{eq:ons_ah3}).}
\label{fig:eff}
\end{center}
\end{figure}

The parameter $ZT$ is readily rewritten
as a function of the sole variable $z$, namely
\be
ZT=\frac{\left( \mbox{sign}(z) \overline L_{ah}(z)-z^2 \overline L_{aa}(z)\right)^2}{\det \overline{L}(z)}.
\ee
The dependence of ZT and $\eta$ on $\tilde h$ are reported in Fig.~\ref{fig:eff} (see the open symbols in
panels (a) and (b)); the solid lines correspond to the low-energy analytic estimates.

$ZT$ diverges in the infinite-temperature limit where, consequently, the efficiency reaches the Carnot limit. 
This behavior is not due to the vanishing of the determinant of $L$ in Eq.~(\ref{eq:zt}), as found for
delta-energy filtering~\cite{mahan96}. 
As it can be easily inferred from Eq.~(\ref{eq:zt}), the divergence of $ZT$ 
is related to: 
{\it i)} the finiteness of $L_{uv}$  and of its determinant (see Fig~\ref{fig:eff}(c)) 
{\it ii)} the divergence of $\mu=m/\beta$.

We can argue that this behavior originates from the fact that infinite-temperature states
are attained for a finite energy density $(\tilde h=2)$.
Let us, in fact, consider the expression of the energy current, 
$j_h \simeq L_{hh}\nabla\beta \simeq -\nabla h$, 
where, for simplicity, we neglect the off-diagonal term due to the gradient of $m=\mu\beta$
and assume a direct proportionality with the energy density gradient $\nabla h$.
From the boundedness of $h$ for vanishing $\beta$, $(2-h)\simeq \beta $ in the C2C model, one obtains $\nabla \beta \sim-\nabla h$. 
Accordingly, a finite current requires $L_{hh}$ to be finite.
Conversely, for standard systems where the energy density diverges with temperature, e.g. $h\sim T^\alpha$ with $\alpha>0$, $\nabla\beta \sim \beta^{\alpha+1} \nabla h$
and a finite current implies a diverging $L_{hh} \sim 1/\beta^{\alpha+1}$.

\section{Conclusions and perspectives}
\label{sec:conc}

In this paper we conducted a fairly detailed study of the transport properties
of a simple one-dimensional model with two conserved quantities: the mass density $a$ and the energy
density $h$.
This model is known to display an equilibrium localization transition when $h$ passes through the critical
value $h_c=2a^2$ and an out-of-equilibrium localization transition if the system is boundary-driven by
suitable reservoirs attached to its ends. 

Because of a scaling relation, the dependence of all thermodynamic variables on $a$ and $h$ can be
reduced to the dependence on a single quantity, typically identifiable with the relative energy
density $h/a^2$.
This includes the Onsager coefficients that we have thoroughly explored in the
homogeneous region, $h \le h_c$.

One of the main outcomes of our study is that a linear-response description of  irreversible transport
processes may apply  even for arbitrarily large temperatures.
Indeed we prove that Onsager coefficients have a smooth
behavior up to the critical curve $T=\infty$, along which they 
exhibit a simple power-law dependence on $m=\mu\beta$. 
Moreover, we have shown that their behavior along the critical line is such that there exists a class of NESS
paths in the $(m,\beta)$ plane that must enter the condensed region.
Negative temperatures are therefore naturally attained by an out-of-equilibrium setup employing reservoirs at positive temperature.
This mechanism suggests a novel effective protocol for the generation of negative-temperature states, which
deserve further studies. In fact, one of the main experimental difficulties 
in this field concerns the ability to thermalize a system at negative temperature
(see~\cite{Baldovin_2021_review} for a review on the topic and~\cite{Baudin23} for a recent experiment).

We have also provided a direct evidence that the Onsager coefficients do depend on nonequilibrium correlations 
and we have identified the largest contribution in correspondence of the critical line of the model. 
Some peculiarities occurring for $T\to\infty$, like the divergence of
the figure of merit $ZT$, are due to the finiteness of Onsager coefficients in such limit,
a property that is strictly related to the finiteness of the energy density when $T$ diverges.

Last but not least, in the low-temperature limit we have obtained an analytic description of the nonequilibrium thermodynamic
observables which compares successfully with the numerical results,
especially if the dynamical rule allows peaks to diffuse. To our knowledge, this is one of the few examples
in which the whole Onsager matrix is exactly computable for an  interacting model.
We have shown that in this limit, spatial correlations are 
absent. Nevertheless, coupled transport is still present, with a positive Seebeck coefficient.

Concerning the perspectives of our work,
we expect that several features observed in the nonequilibrium C2C model
are relevant also for the DNLS equation and its applications.  Two important distinctions,
however, should be emphasized. First of all
no exact scaling properties hold in the DNLS equation because its total energy 
is made of two terms which scale differently with the mass~\cite{Rasmussen2000_PRL}.
The presence of an interaction (hopping) term is particularly relevant at low-temperatures, where we expect
substantial differences between the two models. As an example, the Seebeck coefficient 
was found to change sign in the homogeneous region of the DNLS equation~\cite{iubini12,Iubini2016_NJP}, while in the C2C
model it is always positive, see Fig.~\ref{fig:SD}.
Secondly, the Hamiltonian character of the DNLS dynamics is certainly richer than the stochastic MMC dynamics employed 
for the C2C model.
In particular, we expect that dynamical effects will be relevant in the localized region of the DNLS model, where
the  existence of an adiabatic invariant freezes the macroscopic dynamics when high peaks appear in the system~\cite{PRL_DNLS}.
Because of that, the investigation of NESS profiles and Onsager coefficients in such a region
is computationally very demanding.
There are, however, reasons to think that it would be worth investigating their behavior.

More specifically, it is reasonable to expect that, similarly to the C2C model,
DNLS profiles can cross the critical line when driven by reservoirs in the homogeneous region.
In fact,
in Ref.~\cite{Iubini2017_Entropy} some of us studied the DNLS equation in a nonequilibrium setup
analogous to that used later in Ref.~\cite{gotti22}:
the DNLS chain was attached to a standard reservoir on one boundary and to a dissipator  
on the other boundary. The dissipator, called \textit{sink} in the title, steadily removes mass and energy from the lattice 
and it can be thought of as a boundary condition imposing $a=h=0$.
Within such set-up the system enters the localized region accompanied by a complicated dynamics.

\ack
We thank S Lepri for many useful suggestions on coupled-transport phenomena and related models.
We are also indebted to P C Semenzara for enlightening  discussions on Monte Carlo methods.
P P acknowledges support from the MIUR PRIN 2017 project 201798CZLJ.

\appendix

\section{Perturbative analysis}
\label{app:pert}

In this appendix, we derive the limiting expressions (\ref{eq:zan}),
using Eqs.~(\ref{eq:orig1}-\ref{eq:orig2}) or, more precisely,
\bea
\tilde m(z) &=& \frac{z^2}{2} + \frac{z}{\sqrt{\pi}} \frac{\mathrm{e}^{-z^2/4}}{1 + \mathrm{erf}(z/2) } , \\
\tilde h(z) &=& \frac{z^2}{2} \left( \frac{1}{\tilde m^2(z)} + \frac{1}{\tilde m(z)} \right) .
\eea
In the limit $T\to 0$, $\tilde h \to 1$ and $z\to +\infty$, therefore $\tilde m(z) \simeq z^2/2$ and
$\tilde h(z) \simeq 1 +2/z^2$, i.e.
\be
z = \sqrt{ \frac{2}{\tilde h -1}}, \;\; \mbox{for} \; \tilde h\to 1 .
\ee
In the limit $T\to\infty$, $\tilde h \to 2$, $z\to -\infty$ and calculations are more lengthy.
By accurately expanding the error function,
\be
\mathrm{erf}(z/2) = -1 + \frac{2 \mathrm{e}^{-|z|^2/4}}{|z|\sqrt{\pi}}
\left [1 - \frac{2}{|z|^2} + \frac{12}{|z|^4} - \frac{120}{|z|^6} \right] \; ,
\label{eq.expansion}
\ee
we find that
$\tilde m(z) \simeq -1 + (4/z^2) - (40/z^4)$ and
$\tilde h(z) \simeq 2 - 4/z^2$, i.e.
\be
z = -\frac{2}{\sqrt{2 -\tilde h}}, \;\; \mbox{for} \; \tilde h\to 2 .
\ee

\section{Evolution without pinning}
\label{sec:nopin}

As explained in the main text and sketched in Fig.~\ref{fig:setup},
the new state in each MMC move must be chosen within the
intersection between a sphere and a plane with the constraint
of positive masses; this means either within a full circle or within three disconnected arcs.
In the latter case, there are two selection options: within the same arc as in the original
configuration; within any of the three arcs with equal probability.

Since the three-arcs solution appears when one mass is significantly larger than the other two,
these two options correspond to either pin a peak, or to allow it diffusing.
All of our simulations in the main text have been made following the former option.
This choice originates from the DNLS equation, where peaks
are dynamically pinned~\cite{PRL_DNLS}. Here below we consider the second option as
it helps singling out the role of diffusion at higher temperatures.
\begin{figure}[ht!]
\begin{center}
\includegraphics[width=0.7\textwidth]{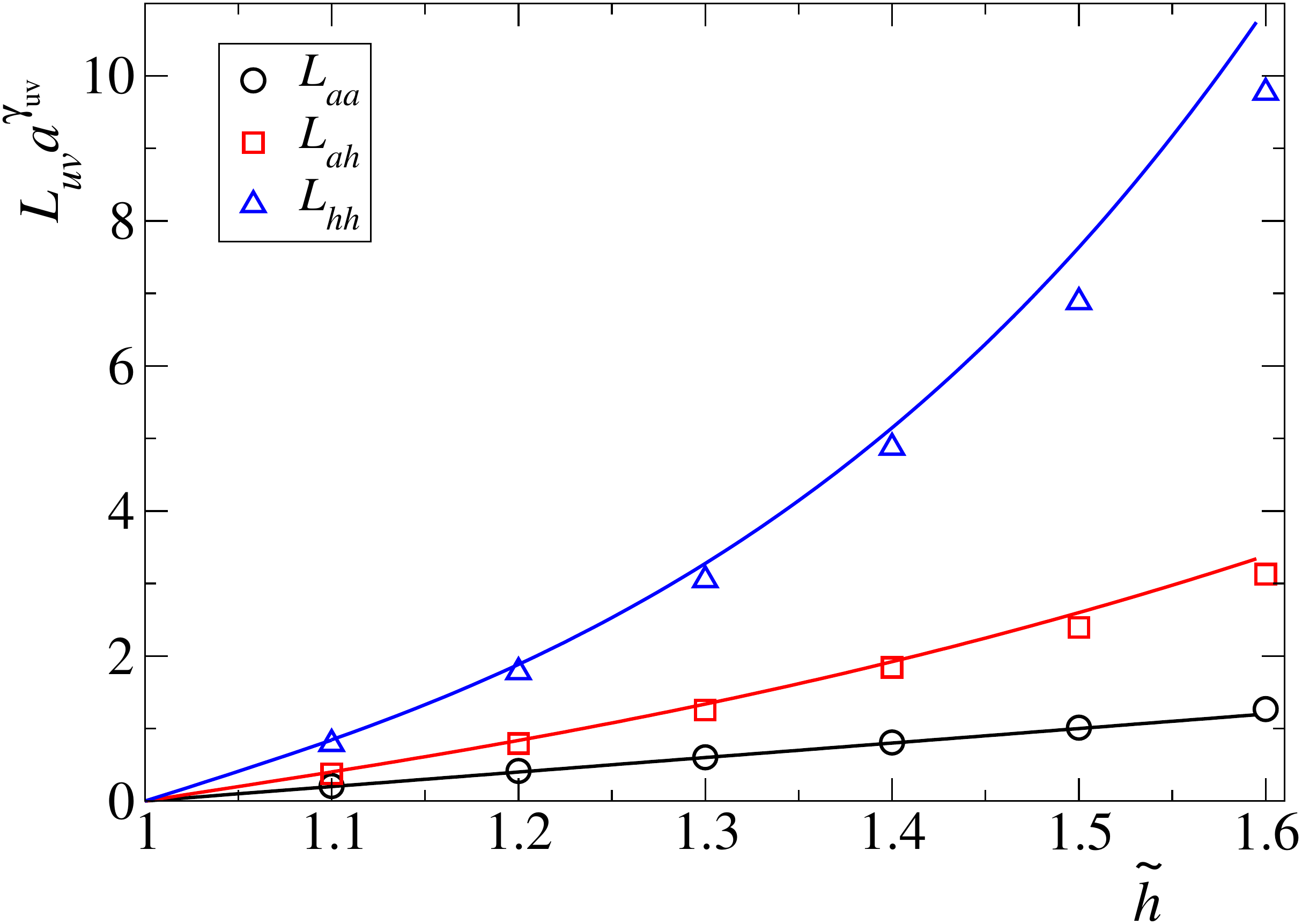}
\caption{
Onsager coefficients versus $\tilde h$ for the unpinned MMC dynamics. Continuous lines are analytic estimates obtained from Eq.~(\ref{eq:ons_ah3}), symbols refer to numerical simulations.  Lower, middle and upper data refer respectively to 
$L_{aa}$, $L_{ah}$ and  $L_{hh}$.
}
\label{fig:l_nopin}
\end{center}
\end{figure}

In fact, when $T$ increases and the three-arcs solution
is increasingly likely, the low-$T$ approximation fails in two respects:
i)~it averages over angles corresponding to unphysical negative masses;
ii)~it allows diffusion	 to the neighboring arcs.
If we adopt a no-pinning evolution, only i) applies. 
In Fig.~\ref{fig:l_nopin} we compare the low-$T$ approximation 
with the numerical outcome of the unpinned model.
The agreement extends to significantly larger energy densities;
in fact, allowing peaks to diffuse, the Onsager (transport) coefficients are now
significantly larger.

\section*{References}

\providecommand{\newblock}{}

\end{document}